%% file: LesHouches_Ecology.tex
\documentclass[submission, PhysLectNotes]{SciPost}

\usepackage{xcolor}
\usepackage{hyperref}
\hypersetup{
    colorlinks=true,
    linkcolor=blue,
    filecolor=magenta,      
    urlcolor=cyan,
}

\hypersetup{
    colorlinks,
    linkcolor={red!50!black},
    citecolor={blue!50!black},
    urlcolor={blue!80!black}
}

\usepackage[bitstream-charter]{mathdesign}
\DeclareSymbolFont{usualmathcal}{OMS}{cmsy}{m}{n}
\DeclareSymbolFontAlphabet{\mathcal}{usualmathcal}

\usepackage{graphicx}
\usepackage{dcolumn}
\usepackage{bm}
\usepackage{dsfont}

\def\be{\begin{equation}} 
\def\ee{\end{equation}} 
\newcommand \bea {\begin{eqnarray}} 
\newcommand \eea {\end{eqnarray}} 
\def \>{\rangle} 
\def \<{\langle}

\begin{document}

\begin{center}{\Large \textbf{
Les Houches Lectures on Community Ecology: From Niche Theory to Statistical Mechanics\\
}}\end{center}

\begin{center}
Wenping Cui \textsuperscript{1},
Robert Marsland III \textsuperscript{2}, and
Pankaj Mehta  \textsuperscript{3$\star$}
\end{center}

\begin{center}
{\bf 1} Kavli Institute for Theoretical Physics, University of California Santa Barbara
\\
{\bf 2} Pontifical University of the Holy Cross
\\
{\bf 3} Dept. of Physics and Faculty of Computing and Data Science, Boston University
\\
${}^\star$ {\small \sf pankajm@bu.edu}
\end{center}

\begin{center}
\today
\end{center}


\section*{Abstract}
{\bf
Ecosystems are among the most interesting and well-studied examples of self-organized complex systems. Community ecology, the study of how species interact with each other and the environment, has a rich tradition. Over the last few years, there has been a growing theoretical and experimental interest in these problems from the physics and quantitative biology communities. Here, we give an overview of community ecology, highlighting the deep connections between ecology and statistical physics. We start by introducing the two classes of mathematical models that have served as the workhorses of community ecology: Consumer Resource Models (CRM) and the generalized Lotka-Volterra models (GLV). We place a special emphasis on graphical methods and general principles. We then review recent works showing a deep and surprising connection between ecological dynamics and constrained optimization. We then shift our focus by analyzing these same models in ``high-dimensions'' (i.e. in the limit where the number of species and resources in the ecosystem becomes large) and discuss how such complex ecosystems can be analyzed using methods from the statistical physics of disordered systems such as the cavity method and Random Matrix Theory.

}

\vspace{10pt}
\noindent\rule{\textwidth}{1pt}
\tableofcontents\thispagestyle{fancy}
\noindent\rule{\textwidth}{1pt}
\vspace{10pt}

\input{1_chapter_one.tex}

\input{2_chapter_two.tex}

\input{4_chapter_four.tex}

\begin{appendix}
\input{5_appendix.tex}

\end{appendix}



\bibliography{references.bib}

\end{document}

%% file: 1_chapter_one.tex
\section{Introduction}
\label{ch:intro}

Life is everywhere on our planet. No matter where one looks, from the boiling hot springs of Yellowstone to the glaciers of Antartica, one can find thriving ecosystems. Despite the enormous variation in temperature, energy supplies, and physical resources, in all these settings organisms interact with each other and their environments to self-organize and survive. Understanding this process is one of the fundamental goals of ecology.

Every ecosystem is different and shaped by the physical environments in which it functions. The organisms that can be found in the desert, where water is a scarce and precious resource, are very different from the kind of organisms that thrive in the rain forest where water is abundant. At a completely different scale, one finds microbial ecosystems -- complex assemblages of microbes that metabolize and produce small molecules. Furthermore, far from being passive creatures in a fixed environment, organisms almost always transform the environments in which they live. The most dramatic example of such a transformation is the great oxidation event two billion years ago by the photosynthetic {\it cynobacteria}. 

Understanding how ecosystems can assemble and function in light of all this complexity and difference has been a central strand of ecological research for the last century. These efforts have birthed an extremely rich theoretical and experimental tradition aimed at identifying the principles underlying community ecology. A central result of this work is the realization that despite all the innumerable differences between ecosystems -- very different physical environments,  the great heterogeneity in organism physiologies and traits --  there exist a common set of core principles that fundamentally shape community assembly and function. In the language of physics, this is the statement ecosystems are shaped by \emph{universal} processes. 

These lectures introduce and build upon one of the most successful and rich theoretical paradigms for understanding ecosystems: niche-theory. As discussed below, niche-theory emphasizes the central role played by ecological competition and selection in shaping communities. Many of the ideas of niche theory have their origins in seminal papers from the 1960's by Richard Levins and Robert MacArthur that introduced many of the core ideas and mathematical models underlying niche theory \cite{levins1962theory, levins1963theory, MacArthur1967, MacArthur1970}. Since then, niche theory has been further developed and expanded upon by many ecologists giving rise to an extremely sophisticated and coherent framework for understanding community ecology. These ideas have been summarized in a number of wonderful books \cite{levins1968evolution, tilman1982resource,chase2003ecological}. For this reason, we are neither capable of, nor have we attempted to, give a comprehensive introduction to the ecological literature.

Instead, the focus of these lectures is on reformulating these ideas with the purpose of presenting them in a language and style more accessible to physicists. This process has often resulted in alternative derivations/formulations of key ideas. Throughout the review, we have tried our best to retain what we consider the best aspects of the theoretical ecology tradition - an emphasis on the use of graphical methods to produce robust theorems using realistic and generalizable models (see Levins' wonderful essay ``The role of Model Building in Population Biology''  \cite{levins1966strategy}) -- and combine this with the best of the physics pedagogical tradition. 

The timing of the lectures is inspired by the exciting and fruitful developments over the last decade in the physics of living systems. This period has seen a renewed interest in niche-theory and community ecology from the viewpoint of statistical physics. These works have sought to extend and integrate many of the ideas of niche theory with techniques and tools from statistical physics, leading to rapid progress and a plethora of new insights.
However, this literature is often scattered across journals in many different fields ( ecology, physics, microbial ecology,  biology) making it difficult to formulate a coherent picture. For this reason we have made a concerted attempt to provide a concise yet coherent introduction to these ideas.

This renewed interest in ecology among physicists has been driven by big experimental advances. This has been especially true of in the microbial realm, where the adoption of quantitative experimental designs centered around the rapid progress in DNA sequencing has allowed for the ability to catalogue the species present in a microbial ecosystem on a scale and resolution that was unimaginable just twenty years ago. This wealth of data has revealed that microbial communities are extremely diverse with tens to thousands of different strains/species present in most ecosystems. This has raised a number of new and fundamental questions about the diversity, function, and structure of microbial ecosystems. To answers these questions,  physicists and ecologists have also started to design quantitative experiments that probe how the physical and chemical environment shapes the structure and function of microbial communities \cite{frentz2015strongly, momeni2017lotka, ratzke2018modifying, Goldford2018,mickalide2019higher, estrela2022functional, dal2021resource, diaz2022top, hu2022emergent, gowda2022genomic}. One of the fundamental insights that has resulted from these experiments are that microbes fundamentally transform the environments in which they live -- suggesting that any theory of microbial ecosystems must account for environmental engineering and the strong coupling between organism and environment \cite{levins1985dialectical, lewontin1997organism}.

The immense diversity seen in real ecosystems combined with the success of alternative frameworks for understanding diversity seen in ecosystems such as neutral theory (see  \cite{rosindell2011unified}  and \cite{azaele2016statistical} for review directed at statistical physicists) has also resulted in a renewed interest understanding how niche theory can be adapted to diverse ecosystems with many species and resources. This has led to a rich body of recent works exploring niche-theory in the ``high-dimensional'' limit. Much of this work draws inspiration from the rich literature in the statistical physics of disordered systems and random matrix theory. 

Unfortunately, there are many interesting recent research directions that we have been unable to cover here. These include new graphical methods that build upon those presented here \cite{wang2021complementary,PhysRevE.108.044409} , the very impressive new work on dynamics and chaos in high-dimensional ecosystems (including dynamical mean-field theories) \cite{pearce2020stabilization,roy2021glassyLV,roy2020persistent,  ros2023rGLVtypical,ros2023rGLVquenched,pirey2023aging, mahadevan2023spatiotemporal},  the extension of consumer resource models to the microbial ecosystems by including metabolism and metabolic cross-feeding \cite{marsland2018available, Goldford2018, marsland2020minimal, mehta2021cross, gowda2022genomic, ho2022competition}, and macroecological laws \cite{grilli2020macroecological}. It is our hope that these lectures can serve as a starting point for readers interested in exploring this literature.

Finally, we note in line with the tradition of the Les Houches lectures we have kept citations to a minimum. We apologize for the many works we have inevitably failed to cite. We wish to emphasize that these lectures reflect the  \emph{personal perspective} of the authors (Cui, Marsland, Mehta), developed over the last few years working in ecology. In no way should these lectures be taken to be a comprehensive or complete overview of the field.  In line with this, many of the derivations of classic results presented in the low-dimensional chapter are new to the literature. However, we have found these derivations to be more intuitive and less mathematical and for this reason have included these in the lectures. 

\begin{figure} 
    \centering
    \includegraphics[width=0.75\textwidth]{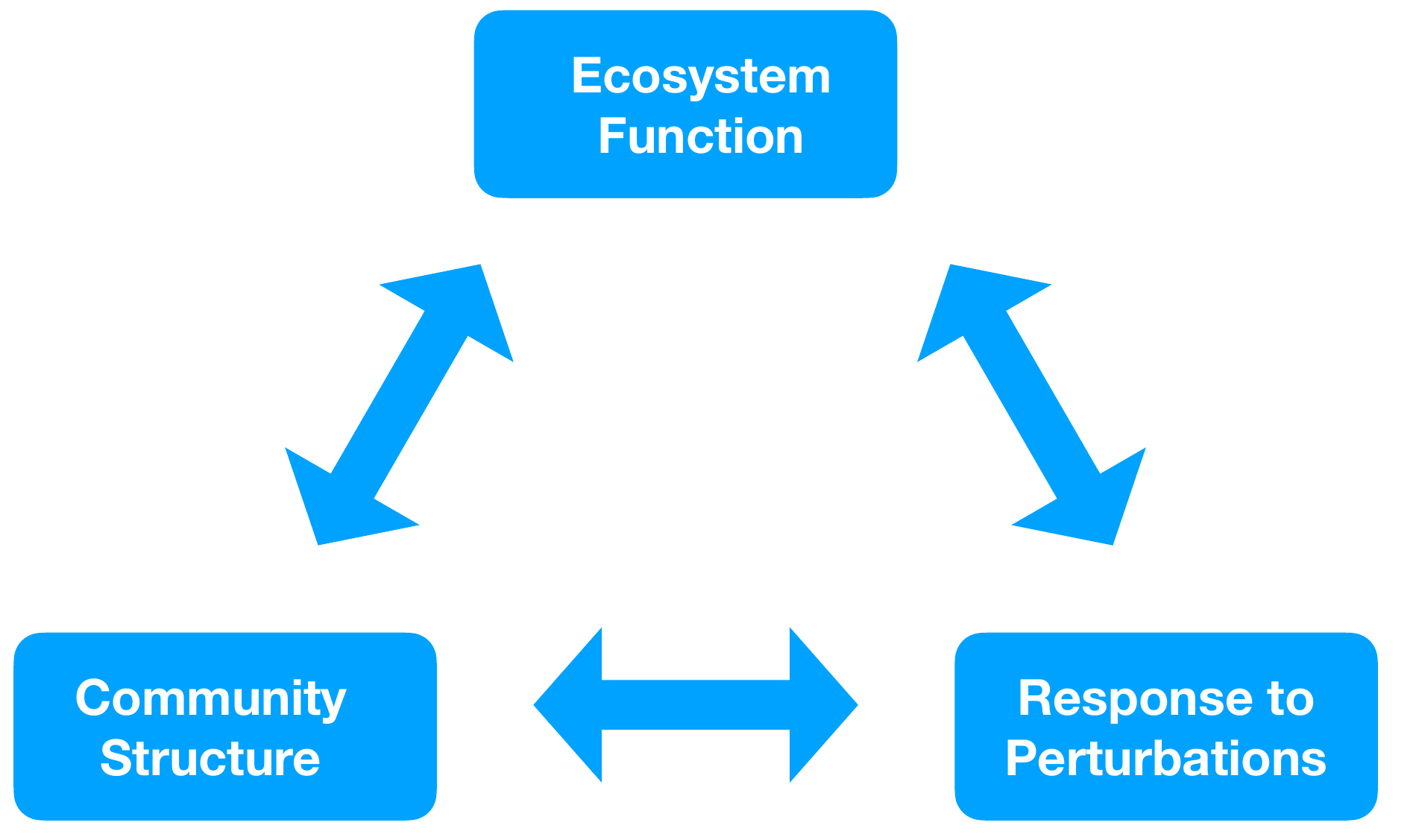}
    \caption{The three core problems and their interrelations.}
    \label{fig:questions}
\end{figure}

\section{Fundamental questions}

Many works in community ecology are concerned with three inter-related problems. The first of these is to understand and predict \emph{community structure}. Community structure is a very broad concept that refers to any property of the communities that reside in an ecosystem \cite{chesson2000mechanisms,ricklefs2004comprehensive,vellend_conceptual_2010,hillerislambers2012rethinking}. These include the \emph{richness} of an ecosystem --the number of different species in a community , the \emph{diversity} of a community -- measures of species biodiversity 
that also accounts for differences in species abundances (e.g. Shannon entropy of species abundances in an ecosystem), and even the \emph{predicability} 
of community assembly - will you end up with the same community in similar environments?

The second major problem in community ecology is to understand \emph{ecosystem function} \cite{tilman1999ecological,downing2002ecosystem,dodson2000relationship,hooper2005effects}.  Ecosystem function is usually defined as the physiochemical and biological processes that occur in an ecosystem that are necessary for maintaining life \cite{srivastava2005biodiversity}. Examples including processes that create biomass and energy, material transformations necessary for ecosystems to thrive (e.g nutrient recycling, biomass production), and often is used to to denote just about any property of an ecosystem that can help sustain human life. This vagueness in definition often leads to numerous debates, controversies, and disagreements. Chief among these is whether the notion of function is something imposed externally by humans or is it an internal property of the ecosystem itself. In our experience, in the literature both of these meanings of ecosystem function are commonly employed.

The third major reoccurring problem addressed in community ecology is understanding how ecosystem properties (including community structure and ecological function)  respond to perturbations \cite{ives2007stability,mccann2000diversity}. Examples of the kind of perturbations that are of interest include changes to the physical environment (e.g. warming of temperatures), the response of the ecosystem to immigration/invasion by new organisms, or the effect temporal disturbances. These questions are often motivated by important practical questions related to the preservation and management of environments or human health (e.g. the effects of antibiotics on the gut \emph{microbiome} - the collection of microbes that live inside the human gut).

A major goal of modern ecological research is to relate these three concepts \cite{loreau2010populations}.
For example, a large line of recent research has been concerned with asking how biodiversity affects ecological function \cite{srivastava2005biodiversity, vellend2017biodiversity}. Another interesting line of research seeks to understand how the structure of food webs shapes community structure and stability in response to perturbations \cite{rooney2012integrating}. 

Finally, we note that while many of these questions and ideas were originally formulated for macroscopic ecosystems such as forests or coral reefs, the last few years have seen an increased interest in using these ideas to understand microbial ecosystems \cite{shade2012fundamentals, hulot2000functional,adler2007niche,chase2019species}. Microbes offer a particularly nice and tractable setting for exploring these problems since it is possible to directly measure community structure using DNA sequencing and the small size of these ecosystems makes it possible to tightly control the physical and biochemical environments.

\section{Ecological Processes}

In order to situate niche theory in the broader context of community ecology,  it is worth briefly considering a very general model of community ecology that incorporates the four major ecological processes thought to be important for ecosystems: (i) \emph{ecological selection} - the idea that organisms that are more suited to an environment will survive in an ecosystem; (ii) \emph{stochasticity} - processes such as demographic noise that introduce randomness; (iii) \emph{dispersal} - species live in local communities and can immigrate between these communities; and finally (iv) \emph{speciation} - the process through which new species can emerge from evolution.  In what follows, we will always focus on timescales over which speciation is negligible (although it can be included as a special kind of ``dispersal'' from an imaginary pool of possible phenotypes).

\begin{figure*}
    \centering
    \includegraphics[width=0.7\textwidth]{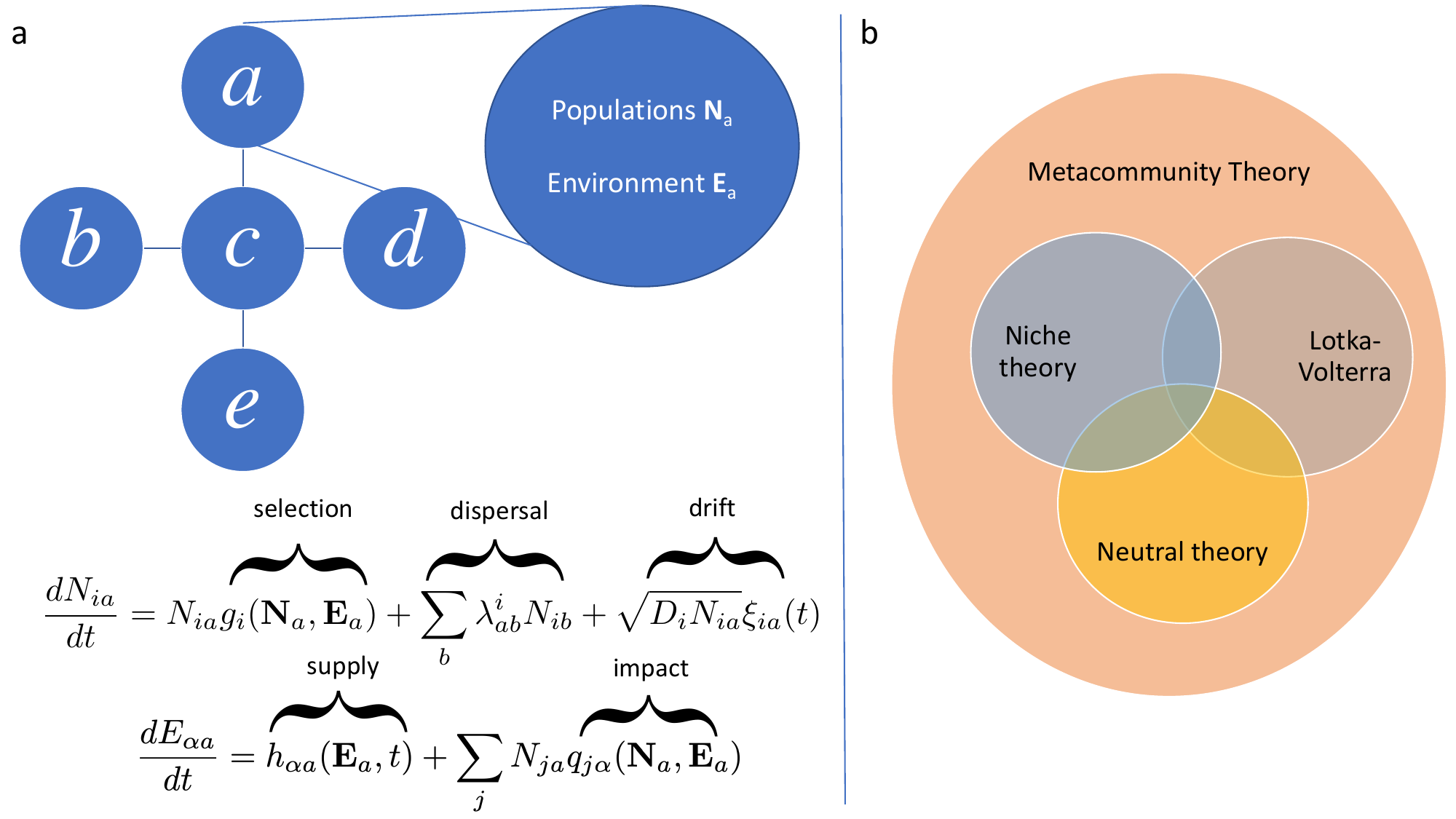}
    \caption{{\bf The metacommunity framework.} \emph{(a)} The metacommunity framework incorporates the three fundamental ecological processes of selection, dispersal and drift within a unified mathematical framework. \emph{(b)} Within this framework, three areas of intense study have been niche theory, generalized Lotka-Volterra theory and neutral theory. The domains of applicability for these theories partially overlap.}
    \label{fig:models}
\end{figure*}

A very general set of  stylized equations that incorporates the first three of these processes is illustrated in Fig. \ref{fig:models}\emph{(a)} and  consists of a set of local patches, $a,b,c,\dots$, a set of species labeled by latin letters $i$ whose population in patch $a$ is given by $N_{ia}$, and  local environmental variables labeled by greek letters $\alpha$ whose value in patch $a$ is given by $E_{\alpha a}$. Within each patch, the dynamics are described by the following set of differential equations:
\begin{align}
\frac{dN_{ia}}{dt} &= N_{ia} g_i (\mathbf{N}_a,\mathbf{E}_a)+\sum_b \lambda^i_{ab}N_{ib} + \sqrt{D_i N_{ia}}\xi_{ia}(t)\label{eq:Ngen}\\
\frac{dE_{\alpha a}}{dt} &= h_{\alpha a}(\mathbf{E}_a,t) + \sum_j N_{ja} q_{j\alpha}(\mathbf{N}_a,\mathbf{E}_a) \label{eq:Egen}
\end{align}
where $\lambda^i_{ab}$ is the migration rate of species $i$ from patch $b$ to patch $a$, and $\xi_{ia}(t)$ is Gaussian white noise. 

The reader should not dwell too much on the details of Equation (\ref{eq:Egen}). This equation is too general to be amenable to mathematical analysis. Instead, the goal of writing Eq. (\ref{eq:Egen}) is to get a sense of how one can represent the ecological processes discussed above mathematically. Notice that the $g_i$ contain the information about ecological \emph{selection} since these functions encode how well a species will grow. The dependence of $g_i$ on $\mathbf{N}$ and $\mathbf{E}$ is the mathematical manifestation of the fact that ecological selection is inseparable from interactions with other species and the shared environment. The $\lambda^i_{ab}$ term represents \emph{dispersal}. This terms controls how species immigrate between patches.  The noise term $\sqrt{D_i N_i}\xi_i(t)$ models the effects of demographic \emph{stochasticity}.  The vector $h_{\alpha a}$ represents the intrinsic behavior of the environment in the absence of other organisms, and is known as the \emph{supply vector} in contemporary niche theory, where it usually refers to an external supply of consumable resources. The impact of the organisms on the environment is captured by the second term, where the per-capita impact of each species $j$ on the environment is measured by an \emph{impact vector} $q_{j\alpha}$.  This impact vector is the mathematical manifestation of the fact that species fundamentally modify the environments in the which they live.

The major branches of ecological theory shown in Fig. \ref{fig:models} each emphasize different aspects of this framework. \emph{Consumer Resource Models} (CRMs) - the main model underlying contemporary niche theory -- and generalized \emph{Lotka-Volterra} (GLV) models -- the main models underlying modern coexistence theory --  focus on interactions with the environment and with other species, respectively. \emph{Neutral theory} considers a different limit, where selective differences among species are absent, and species distributions are driven by drift and dispersal. Despite the differences of focus, these models have some overlap: interactions with the environment in CRMs produce effective interactions between species and thus can be implicitly included in a GLV model. Both of these models reduce to neutral theory in the limit of vanishing selective differences ($g_i = g$ for all $i$).  Recent work from statistical physics literature also suggests that niche and neutral theory are actually better thought of as theories for two different ``ecological phases''  -- akin to a ferromagnet and paramagnet in spin systems -- rather than complete theories themselves \cite{fisher2014transition, Kessler2015}.

In the following sections, we give an overview of niche theory. Niche theory neglect drifts and dispersal and focus on the effects of selection. To build ecological intuitions, we start by considering ``low-dimensional'' ecosystems with 2-3 species and 2 environmental factors. We focus on the two most common types of mathematical models: generalized consumer resource models (CRMs) and generalized Lotka-Volterra models (GLVs). Whereas CRMs explicitly model the environment, GLVs model the environment implicitly through its effect on species-species interactions.
Having build our ecological intuition in this low-dimensional setting, we then consider ``high-dimensional'' ecology where both the number of species and resources are assumed to be large. In this setting, we show how ideas from the statistical physics of disordered systems can be adapted to ecology in order to understand complex ecosystems.

%% file: 2_chapter_two.tex
\label{ch:nichetheory}

\section{An introduction to consumer resource models}
\label{sec:niche}

One of the most influential class of ecological models in theoretical ecology are consumer resource models (CRMs). CRMs assume that all interactions between species are mediated by the consumption of external resources. Mathematically, this means that the $g_i$ and $q_{j \alpha}$ in Eq. (\ref{eq:Ngen}-\ref{eq:Egen}) are functions of $\mathbf{E}$ alone, and do not depend directly on the population sizes $\mathbf{N}$. The only environmental factors considered in CRMs are consumable resources. For this reason, we use the more familiar notation $R_\alpha$ instead of $E_\alpha$, where each $R_\alpha$ represents the abundance of one type of resource in the local patch. As noted above, CRMs neglect drift and dispersal focusing on selection. With these restrictions, we obtain the following set of differential equations for the dynamics within a single patch:
\begin{align}
\frac{dN_i}{dt} &= N_i g_i(\mathbf{R})\label{eq:nicheN}\\
\frac{dR_\alpha}{dt} &= h_\alpha(\mathbf{R}) + \sum_j N_j q_{j\alpha}(\mathbf{R}).
\label{eq:nicheR}
\end{align}

\begin{figure*}
    \centering
    \includegraphics[width=0.78\textwidth]{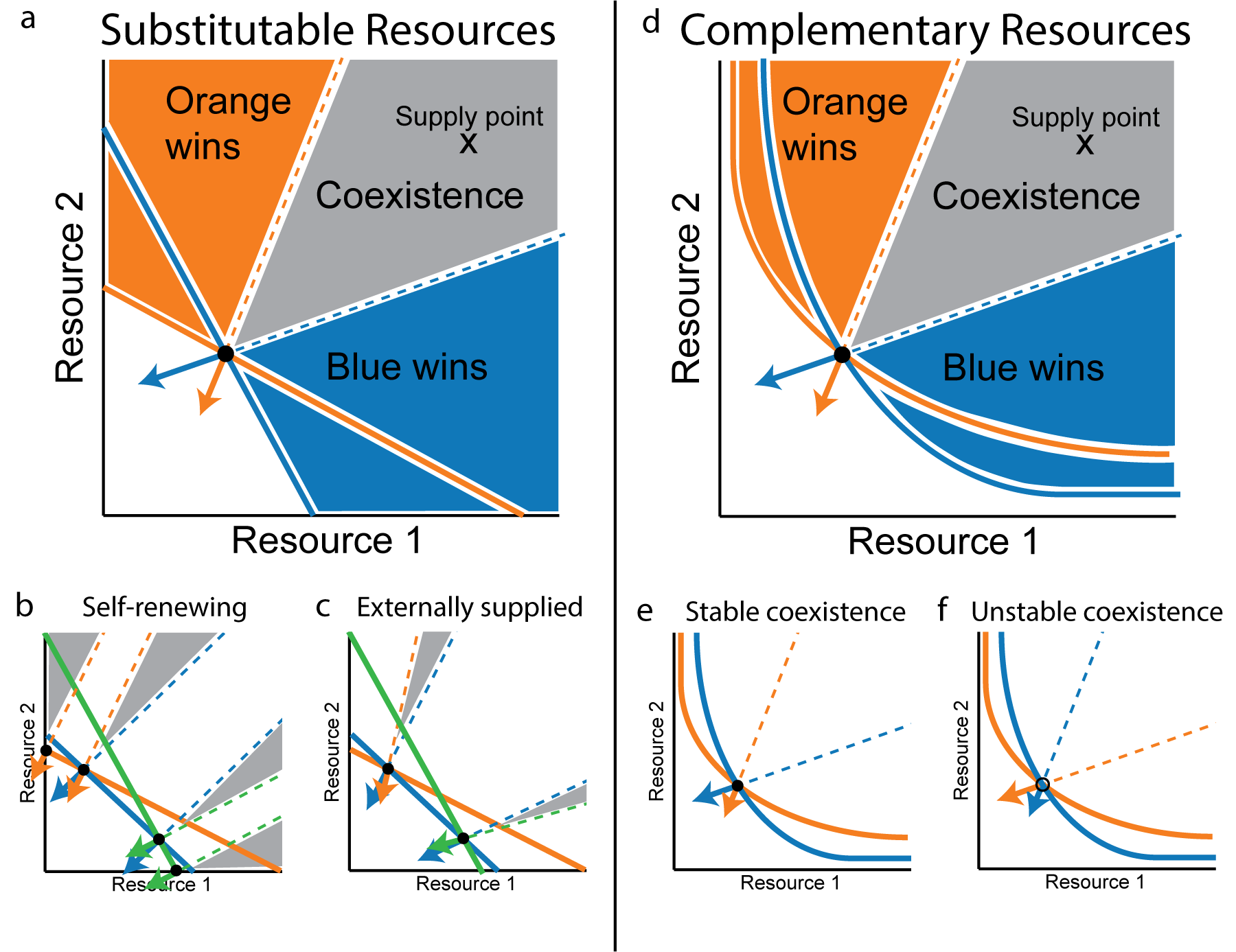}
    \caption{{\bf Niche Theory Examples.} \emph{(a,b,c)} Different resource supply choices with substitutable resources yield different coexistence cone geometries. \emph{(d,e,f)} Stability depends on relative orientation of impact vectors.}
    \label{fig:niche}
\end{figure*}

The elimination of direct interactions between species gives this class of models a bipartite structure, which is amenable to graphical representation, especially for ecosystem with two resources. Fig. \ref{fig:niche} shows how the growth rate, impact vectors and supply can be represented in the two-dimensional space of resource abundances. The set of environmental states $\mathbf{R}$ where $g_i(\mathbf{R})=0$ comprise the zero net growth isocline (ZNGI), a line that separates states of negative growth rate from states of positive growth rate for each species. The impact $q_{i\alpha}(\mathbf{R})$ for each species can be represented as a vector field in this space, as can the supply vector $h_\alpha(\mathbf{R})$. 

This representation immediately leads to geometric criteria for identifying fixed points. In two dimensions, there is an established set of three geometric conditions that together guarantee stable coexistence of two species in a large class of models \cite{tilman1980resources,tilman1982resource,chase2003ecological}:
\begin{enumerate}
    \item The ZNGI's of both species must intersect.
    \item The supply vector $h_\alpha$ (evaluated at the intersection point) must lie within the cone spanned by the negative impact vectors $-q_{i\alpha}$ (also evaluated at the intersection point).
    \item The impact of each species must be biased relatively more towards the resource that has a relatively larger effect on its own growth rate.
\end{enumerate}  
The first condition immediately follows from requiring $dN_i/dt = 0$ for all surviving species, and is valid in any dimension. This is the source of the \emph{Competitive Exclusion Principle}, which says that at most $M$ consumer species can stably coexist.  The competitive exclusion principle is simply the statement that, generically, at most $M$ ZGNIs can intersect at a single point in an $M$-dimensional space in the absence of additional constraints or degeneracies in growth rates. 

The second condition guarantees that a set of positive population sizes $N_i$ can be found whose combined impact $\sum_j N_j q_{j\alpha}$ perfectly balances the supply $h_\alpha$, so that $dR_\alpha/dt = 0$. Rewriting the second equation in Eq.~\ref{eq:nicheR} with the left hand side set to zero gives
\be
h_\alpha(\mathbf{R}) = \sum_j N_j \left( - q_{j\alpha}(\mathbf{R}) \right).
\ee
Since species abundances must be positive $N_j>0$, the equation has the geometric interpretation that the the supply $h_\alpha$ must lie within the cone spanned by the negative impact vectors.
This condition is also valid in any dimension, and for any number of species, not just in $M=S=2$. For the most commonly used models the supply vector can be regarded as pointing directly towards a ``supply point,'' as depicted in Fig. \ref{fig:niche}.  The supply point represents the steady-state resource abundances in the absence of consumers. The graphical representation is further simplified in such cases, since the full supply vector field no longer needs to be directly considered, and the satisfaction of the coexistence condition depends simply on whether the supply point (which is often modeled as a directly accessible control parameter) falls inside the coexistence cone.

The third condition is the most subtle. The first two conditions establish that $dN_i/dt = dR_\alpha/dt = 0$, but imply nothing about the stability of this fixed point. In two dimensions, Tilman showed that the stability of the fixed point in a large class of two-dimensional models can be determined by comparing the relative orientation of the impact vectors to the relative orientation of the ZNGI's (see also Appendix \ref{app:limit}) \cite{tilman1980resources}.

In the rest of this section, we illustrate this graphical approach and the conclusions that can be drawn from it using three canonical examples. See Appendix \ref{sec:crm} for a systematic taxonomy of the main kinds of classic niche theory models, with additional examples.

\subsection{MacArthur's Consumer Resource Model}
We begin with MacArthur's Consumer Resource Model (MCRM), which was the first model developed within this framework, and remains a fundamental reference point for ecological theory (\cite{MacArthur1970,chesson_macarthurs_1990}). This model describes competition for ``substitutable'' resources using simple mass-action kinetics. Individuals of consumer species $i$ encounter and consume units of resource $\alpha$ at a rate $c_{i\alpha} R_\alpha$, where the $c_{i\alpha}$ are constants expressing the preferences of species $i$ for the different types. Resource $\alpha$ carries an energy density $w_\alpha$. Some amount of energy per individual $m_i$ is required to sustain the population at steady state, and excess energy leads to population growth, with an efficiency factor $e_i$ converting from energy intake to growth rate. Finally, the resources themselves are self-renewing, such that in the absence of consumers, each resource type $\alpha$ follows its own independent logistic growth law with carrying capacity $K_\alpha$ and low-density growth rate $r_\alpha$. Putting this all together, we have:
\begin{align}
\frac{dN_i}{dt} &= N_i e_i \left[ \sum_{\beta} w_\beta c_{i\beta} R_\beta - m_i\right]\label{eq:MacN}\\
\frac{dR_\alpha}{dt} &= \frac{r_\alpha}{K_\alpha} R_\alpha(K_\alpha - R_\alpha) - \sum_{j} c_{j\alpha} N_j R_\alpha 
\label{eq:MacR}.
\end{align}

Graphical representations of this model for two or three species and two resources are shown in Fig. \ref{fig:niche}\emph{(a,b)}. In sketching these figures, we have used the fact that the equilibrium condition $dR_\alpha/dt = 0$ for each resource can be arbitrarily rescaled without affecting the results of the analysis. The standard practice in the literature is to multiply both sides by $w_\alpha/R_\alpha$, yielding:
\begin{align}
    0 &= {w_\alpha r_\alpha}{K_\alpha}(K_\alpha - R_\alpha) - \sum_j w_\alpha c_{j\alpha} N_j.
\end{align}
This can equivalently be viewed as the equilibrium condition for a model with supply vector
\begin{align}
    h_\alpha = {w_\alpha r_\alpha}{K_\alpha}(K_\alpha - R_\alpha)
\end{align}
which points directly at the supply point $K_\alpha$, and impact vector
\begin{align}
    q_{i\alpha} = -w_\alpha c_{i\alpha}
\end{align}
which is minus the gradient $\partial g_i/\partial R_\alpha$ of the growth rate. This rescaling is  extremely convenient for the graphical representation because the impact vectors are always perpendicular to the ZNGI's, and the coexistence condition reduces to the requirement that the supply point fall inside the (negative) cone spanned by these vectors. If the supply point falls outside this cone but above at least one of the ZNGI's, then one of the species will persist, driving the other one extinct. As shown in Fig. \ref{fig:niche}\emph{(a)}, this allows us to divide up the whole resource space into three regions based on the outcome of competition when the supply point is placed within the region: the ``coexistence cone'' where both species persist and two regions where one of the species competitively excludes the other. Note that the third criterion for stable coexistence above is always satisfied in this model, because the angle of the impact vector is tied to the angle of the ZNGI, and so the equilibrium point is always linearly stable. 

Fig. \ref{fig:niche}\emph{(b)} illustrates some additional features of this model that follow from the graphical representation. In the figure, we have drawn ZNGI's for three species, resulting in two possible coexistence cones. Notice as expected from the competitive exclusion principle, at most two of the species can coexist at anytime. The first thing to note is that the third intersection, between the orange and green species, is not a possible coexistence point, because the blue species has a positive growth rate there, and will always re-invade. The second feature is the presence of two cones at the intersection of the inner ZNGI's with the resource axes. The existence of these cones does not depend on the number of species, but they were omitted from panel \emph{(a)} for simplicity, since they do not affect the outcome of competition. When the supply point lies inside one of these cones, one of the resources is driven to extinction, and the equilibrium state lies at the intersection of the ZNGI with the resource axis. 

When these cones are included in the diagram, then the competitive exclusion limit is saturated whenever the supply point falls in any cone. Although only one species survives in the boundary cones, the number of species still equals the number of \emph{persisting} resources. This leads us to the final interesting feature of the diagram, which is the fact that the inner boundaries of neighboring cones are always parallel. This follows immediately from the fact that the impact vectors are perpendicular to the ZNGI's. It implies that the cones cover the full 90-degree angle of the quadrant of positive resource abundances. As the magnitude of the supply grows (i.e., the distance of the supply point from the origin), it therefore requires increasingly precise fine-tuning of the supply point to avoid landing in one of the cones, with the result that the competitive exclusion bound is generically saturated in the limit of large supply.

\subsection{Externally supplied resources}
The logistic growth of the resources in the absence of consumers makes the MCRM a reasonable model of competition for substitutable biotic resources, such as for herbivores competing for plants. But if the resources are externally supplied nutrients, a different model is needed. For example, in microbial ecology, there can be competition for various organic molecules that can be used as carbon sources. In such cases, a better model for the supply vector is a simple linear law $h_\alpha = \kappa_\alpha - \omega_\alpha R_\alpha$ such as would arise from an ideal chemostat with incoming nutrient flux $\kappa_\alpha$ and dilution rate $\omega_\alpha$. With this supply, we obtain:
\begin{align}
\frac{dN_i}{dt} &= N_i e_i \left[ \sum_{\beta} w_\beta c_{i\beta} R_\beta - m_i\right]\label{eq:extN}\\
\frac{dR_\alpha}{dt} &= \kappa_\alpha -\omega_\alpha R_\alpha - \sum_{j} c_{j\alpha} N_j R_\alpha.\label{eq:extR}
\end{align}
In this case, the true supply vector already points directly towards the supply point $\kappa_\alpha/\omega_\alpha$. Notice that the impact vector $q_{i\alpha} = -c_{i\alpha}R_\alpha$ depends on the resource concentration and is no longer perpendicular to the ZNGI. The factor of $R_\alpha$ rotates it towards the most abundant resource, compressing all the supply cones towards the resource axes as shown in Fig. \ref{fig:niche}\emph{(c)}. 

The first consequence of this modified cone geometry is that the extinct resource cones disappear. When $R_\alpha = 0$ for one of the resources, the impact vector becomes parallel with the other resource axis, shrinking the angle spanned by this cone to zero. The fact that no equilibrium state can have $R_\alpha = 0$ for any $\alpha$ can also be seen by direct inspection of the Eq. (\ref{eq:extR}), where $dR_\alpha/dt = \kappa_\alpha >0$ at zero concentration of resource $\alpha$. The second consequence is that adjacent coexistence cones no longer have parallel boundaries. This means that a finite fraction of possible supply points remain outside all the cones even in the limit of infinite total supply. In the high-dimensional setting, this prevents random high-dimensional ecosystems with linear resource dynamics from saturating the competitive exclusion bound \cite{cui2019effect}.

\subsection{Essential resources}
Both of the previous examples assumed that resources were substitutable, and that the same growth rate could be obtained with any combination of resources with the same combined energy content. But many of the resources required by organisms play distinct functional roles, all of which must be filled for the organism to survive. Plants, for example, require sufficient supplies of carbon, nitrogen and phosphorous to be present in the soil. If any one of the three is too scarce, the plant will not survive, regardless of the available quantities of the other two. This phenomenon implies that the ZNGI's do not intersect with any of the axes, and must have the roughly hyperbolic shape sketched in Fig. \ref{fig:niche}\emph{(d)}. Such ZNGI's can be obtained from a model of the form:
\begin{align}
\frac{dN_i}{dt} &= N_i \left\{ \left[\sum_{\beta} (c_{i\beta} R_\beta)^n \right]^{\frac{1}{n}} - m_i\right\}\label{eq:essN}\\
\frac{dR_\alpha}{dt} &=\kappa_\alpha -\omega_\alpha R_\alpha- \sum_{j} v_{j\alpha}N_j  \left[\sum_{\beta} (c_{i\beta} R_\beta)^n \right]^{\frac{1}{n}}\label{eq:essR}
\end{align}
with $n<-1$. In the limit $n\to -\infty$, this model reduces to Liebig's Law of the Minimum, where the growth rate of each species only depends on the concentration of the most limiting resource:
\begin{align}
\frac{dN_i}{dt} &= N_i \left[\min_{\beta}(c_{i\beta} R_\beta)-m_i\right]\label{eq:essNl}\\
\frac{dR_\alpha}{dt} &=\kappa_\alpha -\omega_\alpha R_\alpha- \sum_{j} v_{j\alpha}N_j \min_{\beta}(c_{j\beta} R_\beta)\label{eq:essRl}
\end{align}
We have chosen the impact vector in the standard way, in which the nonlinear term $\left[\sum_{\beta} (c_{i\beta} R_\beta)^n \right]^{\frac{1}{n}}$ is treated as a total resource uptake rate, which is partitioned among the available resources according to the fixed biomass stoichiometry of the species $v_{i\alpha}$ \cite{tilman_resource_1982}. 

In this model, the direction of the impact vector is set by the biomass stoichiometry $v_{i\alpha}$, regardless of the resource concentration and of the parameters $c_{i\alpha}$ and $n$ that define the shape of the ZNGI. This means that the third criterion for stable coexistence is not necessarily satisfied, and fixed points can exist that are linearly unstable. Panels \emph{(e)} and \emph{(f)} of Fig. \ref{fig:niche} represent two possible choices of biomass stoichiometry that lead to the same coexistence cone, with only the first one leading to stable coexistence. 

The existence of different kinds of fixed points in this model opens the door to more complex attractors. For example, limit cycles, heteroclinic cycles, and chaos have all been observed in models of this basic type when $M,S \geq 3$ \cite{huisman1999biodiversity, huisman2001biological, huisman2001fundamental}.

\section{Generalized Lotka-Volterra Theory}
In the introduction to these lectures, we noted that ecological theory considers two kinds of interactions that affect an organism's growth rate: interactions with the environment and interactions with other species. In the previous section, we saw how CRMs focus on the first kind of interaction. Historically, the earliest attempts at mathematical modeling of ecosystems focused on the second kind of interactions, reducing the role of the environment to a mediator of species-species interactions  \cite{volterra_variation_1926,macarthur1967limiting,macarthur1970species}. The most influential model for understanding species-species interactions is the Lotka-Volterra model. This model was first developed by Lotka and Volterra to describe predator-prey oscillations, and later extended to more general scenarios of interactions among larger collections of species \cite{volterra_variation_1926,lotka_contribution_1909, fisher2014transition}. 
 
In this section, we derive the multi-species (or ``generalized'') Lotka-Volterra equations (GLV) as a limit of the framework introduced in eq. (\ref{eq:Ngen}-\ref{eq:Egen}). We then review how these equations have been used to analyze coexistence in two-species communities. Finally, we provide additional intuition for these results and point out potential complications by applying them to the effective interactions that emerge from the consumer resource models (CRMs) studied above. We will see that ``higher-order'' interactions generically arise from resource-explicit models, and that often, the strength of interaction between a pair of species becomes context-dependent. 


\subsection{Deriving the GLV Equations}
From the physicist's perspective, the GLV equations are analogous to elasticity theory. We take a complex nonlinear system and describe it in terms of the phenomenological linear-response parameters that characterize its behavior near a stable fixed point. This can be straightforwardly carried out in the case of a fixed environment $\mathbf{E} = {\rm const}$. Expanding the growth rate $g_i$ to linear order in the perturbation $\mathbf{N}-\bar{\mathbf{N}}$ from a fixed point $\bar{\mathbf{N}}$, we obtain:
\begin{align}
    \frac{dN_i}{dt} = N_i\left[g_i + \sum_j \frac{\partial g_i}{\partial N_j}(N_j - \bar{N}_j) + \mathcal{O}((N_j-\bar{N}_j)^2)\right]
\end{align}
where $g_i$ and its derivatives are evaluated at $\mathbf{N} = \bar{\mathbf{N}}$. If all species are present at nonzero abundance then we know $g_i(\bar{\mathbf{N}})=0$. But the $g_i$ is included in this expression to allow for expansion about fixed points where $N_i = 0$ for some $i$. This will be required in the invasion analysis below. 

If this were the exact model, with no higher-order terms, the intrinsic growth rate $r_i$ of species $i$ would be found by extrapolating to $N_j = 0$ for all $j$ in the sum, yielding:
\begin{align}
    r_i = g_i(\bar{\mathbf{N}})-\sum_j \frac{\partial g_i}{\partial N_j} \bar{N}_j.
\end{align}
The effect of other organisms on the growth rate is given by the ``competition coefficients'' $\alpha_{ij}$, equal to
\begin{align}
    \alpha_{ij} = -\frac{\partial g_i}{\partial N_j}.
\end{align}
In terms of these parameters, the local dynamics are
\begin{align}
    \frac{dN_i}{dt} = N_i\left[r_i - \sum_j \alpha_{ij} N_j+\mathcal{O}((N_j-\bar{N}_j)^2)\right].
    \label{eq:GLV}
\end{align}
This is the canonical GLV equation. Since the derivatives that define $r_i$ and $\alpha_{ij}$ are evaluated at $\mathbf{N}=\bar{\mathbf{N}}$, they will in general take on different values at different fixed points, including when different assemblages of species are present.

The GLV framework is frequently employed to model environmentally mediated interactions as well as direct interactions. Doing this requires somehow writing the environmental state as a function of the population sizes $\mathbf{E}(\mathbf{N})$. Once this is achieved, the growth rate $g_i(\mathbf{N},\mathbf{E}(\mathbf{N}))$ is again a function of the population sizes alone, and the GLV parameters are readily obtained using the chain rule:
\begin{align}
    r_i &= g_i(\bar{\mathbf{N}},\bar{\mathbf{E}})-\sum_j \left(\frac{\partial g_i}{\partial N_j} + \sum_\beta \frac{\partial g_i}{\partial E_\beta}\frac{\partial E_\beta}{\partial N_j}\right) \bar{N}_j\label{eq:ri}\\
    \alpha_{ij} &= -\frac{\partial g_i}{\partial N_j} - \sum_\beta \frac{\partial g_i}{\partial E_\beta}\frac{\partial E_\beta}{\partial N_j}.\label{eq:alphaij}
\end{align}
If the environmental dynamics have a stable fixed point for arbitrary fixed $\mathbf{N}$, this fixed point defines a function $\mathbf{E}(\mathbf{N})$ that can be used in the above definitions. But the resulting GLV dynamics do not necessarily match the original dynamics, even in the vicinity of the fixed point \cite{o2018whence}. 


\subsection{Invasion analysis and Modern Coexistence Theory}
One of the main uses to which the GLV equations have been put is the derivation of conditions for species coexistence. A categorization of the main factors affecting coexistence has been developed with the help of this framework, and is known as Modern Coexistence Theory (MCT) \cite{chesson2000mechanisms,letten2017linking,barabas2018chesson}. Instead of studying the properties of fixed points, as we did with niche theory above, MCT obtains sufficient conditions for coexistence in terms of the ability of each species to invade the rest of the community. While this approach has some limitations, which will emerge in the examples below, it has the distinct advantage of not depending on how well the GLV equations approximate the dynamics. The analysis is conducted entirely in terms of the sign of the growth rate $g_i$ of the hypothetical invader at a fixed point of the resident community, for which the GLV equations provide an exact expression.

The central results of this theory concern the interaction of two species \cite{chesson2000mechanisms,letten2017linking}. We start by computing the equilibrium abundance of each species $i$ when it is the only species present:
\begin{align}
    \bar{N}_{i/j} = \frac{r_{i/j}}{\alpha_{ii/j}}
\end{align}
where the $/j$ in the subscript indicates that these quantities are being evaluated at the fixed point where the other species $j$ is absent. If we now imagine introducing species $j$ at vanishingly small density, its initial per-capita growth rate is:
\begin{align}
    g_{j/j} &= r_{j/j} - \alpha_{ji/j}\bar{N}_{i/j} \nonumber \\
    &= r_{j/j} - \alpha_{ji/j}\frac{r_{i/j}}{\alpha_{ii/j}}.
\end{align}
For species $j$ to successfully invade, this quantity must be positive, which implies:
\begin{align}
    \frac{r_{j/j}}{r_{i/j}} \frac{\alpha_{ii/j}}{\alpha_{ji/j}} > 1.
\end{align}
The two species $i$ and $j$ are guaranteed to stably coexist when each one is capable of invading the other, that is, when:
\begin{align}
    \frac{r_{j/j}}{r_{i/j}} \frac{\alpha_{ii/j}}{\alpha_{ji/j}} > 1 > \frac{r_{j/i}}{r_{i/i}} \frac{\alpha_{ij/i}}{\alpha_{jj/i}}.\label{eq:coexist1}
\end{align}
This is a sufficient condition for coexistence of two species in any model within our framework, as long as the relationship $\mathbf{E}(\mathbf{N})$ can be defined. It is not a necessary condition, because it is possible for a species that fails to invade at low density to successfully invade at high density. But it does facilitate a robust classification of factors that promote coexistence.

This classification is normally formulated by distinguishing two contributions to the expressions that appear in the inequality. This distinction implicitly assumes that the GLV parameters are independent of which species is resident, so that the $/j$ notation can be dropped from eq. (\ref{eq:coexist1}). The quantities $r_{i}, r_{j}$ can then be interpreted as estimates of the intrinsic ``fitness'' of the each species, and the same fitness ratio $r_{j}/r_{i}$ appears on both sides of (\ref{eq:coexist1}). The expression can therefore be rearranged to give:
\begin{align}
    \frac{\alpha_{ii}}{\alpha_{ji}} > \frac{r_i}{r_j} > \frac{\alpha_{ij}}{\alpha_{jj}}\label{eq:coexist}.
\end{align}
Factors that tend to reduce fitness differences (making $r_i/r_j$ closer to 1) are known as ``equalizing mechanisms'' of coexistence. But the coexistence condition (\ref{eq:coexist}) shows that equalizing mechanisms are insufficient to achieve coexistence, because even as the ratio approaches 1, the ratios of competition coefficients still remain. We see that coexistence requires at least one of the species to compete with itself more than it competes with the other species (i.e., $\alpha_{ii}>\alpha_{ij}$). Factors that increase competition among members of the same species relative to competition with the other species are known as ``stabilizing mechanisms'' of coexistence.

In the following sections, we present the GLV approximations to the CRMs presented in Section \ref{sec:niche} above. In the original MCRM (eq. \ref{eq:MacN}-\ref{eq:MacR}), the approximation is exact, and stabilizing mechanisms have a mechanistic interpretation in terms of decreased niche overlap. When the resources are externally supplied (eq. \ref{eq:extN}-\ref{eq:extR}), the GLV parameters become context-dependent, and the equalizing and stabilizing mechanisms can no longer be cleanly separated as in eq. (\ref{eq:coexist}). 

\subsection{MacArthur's Consumer Resource Model}
Additional intuition about the operation of stabilizing mechanisms comes from analyzing the GLV approximation to the MCRM \cite{MacArthur1970,chesson_macarthurs_1990}. We will consider a slightly generalized version of the model, allowing depletion rates $d_{i\alpha}$ to be independent of the growth law parameters $c_{i\alpha}$:
\begin{align}
    \frac{dN_i}{dt} &= N_i\left[ \sum_\beta w_\beta c_{i\beta} R_\beta - m_i\right]\\
    \frac{dR_\alpha}{dt} &= \frac{r_\alpha}{K_\alpha}R_\alpha(K_\alpha - R_\alpha) - \sum_j d_{j\alpha} N_j  R_\alpha.
\end{align}
Following the procedure outlined above, we write the resource abundances as a function of the population sizes by setting $dR_\alpha/dt = 0$. This equation has two solutions, $R_\alpha = 0$, and 
\begin{align}
R_\alpha = K_\alpha\left(1-\frac{1}{r_\alpha}\sum_j d_{j\alpha}N_j\right).
\end{align}
If the term in brackets is negative, then $dR_\alpha/dt < 0$ for all positive $R_\alpha$, and so $R_\alpha = 0$ is the stable solution. Otherwise, $R_\alpha = 0$ is unstable, because a small addition of the resource will lead to a positive growth rate. The stable equilibrium state is therefore given by
\begin{align}
    R_\alpha(\mathbf{N}) &= {\rm max}\left[0, K_\alpha\left(1-\frac{1}{r_\alpha}\sum_j d_{j\alpha}N_j\right)\right].
\end{align}
Inserting this into eqs. (\ref{eq:ri}), we obtain an effective GLV of the form:
\begin{align}
    \frac{dN_i}{dt} = N_i\left[r_i - \sum_j \alpha_{ij} N_j+\mathcal{O}((N_j-N_j^*)^2)\right].
\end{align}
with
\begin{align}
    r_i    &= \sum_{\beta\in \mathbf{M}^*} w_\beta c_{i\beta} K_\beta- m_i \label{eq:riMCRM}
\end{align}
and
\begin{align}
    \alpha_{ij} &= \sum_{\beta\in \mathbf{M}^*} \frac{w_\beta K_\beta}{r_\beta} c_{i\beta}d_{j\beta},
    \label{eq:alphaMCRM}
\end{align}
where $\mathbf{M}^*$ represents the set of resources with nonzero abundance. Eq.~\ref{eq:riMCRM} has a simple interpretation. The intrinsic growth rate $r_i$ is the difference between the maximum consumption capacity $\sum_{\beta\in \mathbf{M}^*} w_\beta c_{i\beta} K_\beta$ at the given resource supply levels and the maintenance cost $m_i$. 

The formula for the competition matrix in eq. (\ref{eq:alphaMCRM}) carries an ecologically significant meaning, showing how the strength of competition is determined by the overlap between the requirements $c_{i\alpha}$ of one species and the impacts $d_{j\alpha}$ of the other. The overlap is weighted by a product of three factors: (a) the value $w_\alpha$ of each resource, with more energy-rich resources counting for more in the sum, (b) the low-density growth timescale $1/r_\alpha$ for each resource, with rapid renewal decreasing the significance of the resource as a locus of competition and (c) the resource carrying capacity $K_\alpha$, giving resources with higher carrying capacity a greater contribution to the interaction.

Note that if no resources are driven to extinction, then both $r_i$ and $\alpha_{ij}$ are independent of which fixed point is chosen (e.g., whether both species, only species $i$ or only species $j$ is present in the two-species case). We can therefore drop the extra index indicating which species is absent, and simply write $r_i$ and $\alpha_{ij}$ as assumed in the simplified form of the coexistence condition (\ref{eq:coexist}).

These considerations suggest that the scaled quantity $\alpha_{ij}/\alpha_{ii}$ is a natural definition of the ``niche overlap'' between species $i$ and $j$. This definition is ambiguous, however, since it is not symmetric under exchange of $i$ and $j$. If we symmetrize by taking the geometric mean, we can define the niche overlap $\rho$ by
\begin{align}
    \rho = \sqrt{\frac{\alpha_{ij}\alpha_{ji}}{\alpha_{ii}\alpha_{jj}}}.
\end{align}
With this definition, and recalling that the values of the GLV parameters do not depend on which species is the resident and which is the invader (as long as no resources go extinct), we can simplify the coexistence condition (\ref{eq:coexist1}) into the more interpretable form \cite{chesson2000mechanisms,letten2017linking}
\begin{align}
    \frac{1}{\rho} > \frac{r_1}{r_2} > \rho. \label{eq:coexist2}
\end{align}
In this context, stabilizing mechanisms are features of the ecosystem that reduce the niche overlap, opening up more space for species with different maintenance costs $m_i$ or consumption capacity $\sum_{\beta\in \mathbf{M}^*} w_\beta c_{i\beta} K_\beta$ (the two components of $r_i$) to stably coexist.

\subsection{Externally supplied resources}
As noted above, a special feature of the MCRM is that the ``intrinsic'' growth rate $r_i$ and the competition coefficient $\alpha_{ij}$ between species $i$ and species $j$ (Eq. \ref{eq:alphaMCRM}) can be fully determined by the properties of these two species and of the resources, regardless of which species or combination of species is currently resident. The one exception occurs when the addition of a third species drives one of the resources extinct, so that the set $\mathbf{M}^*$ of resources to be summed over changes. This caveat is often ignored in the literature. 

In other models, the context-dependence of the GLV parameters is much more apparent. For the case of externally supplied substitutable resources (Eq. \ref{eq:extN}-\ref{eq:extR}), solving for the resource equilibrium yields:
\begin{align}
    R_\alpha(\mathbf{N}) = \frac{\kappa_\alpha}{\omega_\alpha + \sum_j c_{j\alpha} N_j}.
\end{align}
Inserting this into the equation for the consumer dynamics (Eq. \ref{eq:extN}) gives:
\begin{align}
    \frac{dN_i}{dt} &= N_i\left[ \sum_\beta \frac{w_\beta c_{i\beta}\kappa_\beta}{\omega_\beta + \sum_j c_{j\beta} N_j} - m_i\right].
\end{align}
This is no longer of the GLV form. But in the vicinity of the equilibrium state $N_i^*$, we can still use eq. (\ref{eq:ri}-\ref{eq:alphaij}) to locally obtain the GLV parameters
\begin{align}
    r_{i} &= \sum_{\beta}w_\beta c_{i\beta} R_\beta^* \frac{2\kappa_\beta - \omega_\beta R_\beta^*}{\kappa_\beta} - m_i \\
    \alpha_{ij} &= \sum_\beta \frac{w_\beta (R_\beta^*)^2}{\kappa_\beta} c_{i\beta}c_{j\beta}.
\end{align}
The contribution of each resource to the sums in both equations now depends on its equilibrium abundance $R_\alpha^*$. This complicates the above analysis of stabilizing versus equalizing mechanisms, since in general it is no longer true that $r_{i/j}\approx r_{i/i}$. In communities with more than two species, this also generates effective higher-order interactions of all orders, since the interaction between any pair of species is affected by  all the other species with overlapping consumption preferences.

\section{Niche theory and optimization}

Ecological models where species interact reciprocally (i.e. how species $i$ affects species $j$ is identical to how species $j$ affects species $i$) can be recast in the language of optimization.
 This optimization perspective provides a powerful set of analytical tools.  In the papers where MacArthur first presents his Consumer Resource Model, he  already started to investigate the conditions under which this ecological model can be represented in terms of an analogous optimization principle \cite{MacArthur1969,macarthur1970species}. Recent work has extended his original results and identified when such an optimization principle can be found \cite{Marsland2019a}. One of the key lessons from these works is that it is much more natural to optimize over the state of the environment (i.e. the resource abundances) than over population sizes. When we do this, the objective function has a transparent ecological interpretation.

Below, we start with MacArthur's original result, where the optimization is performed over the space of population sizes in the GLV approximation to the MCRM. We then briefly describe the optimization problems in resource space solved by the equilibrium states of the CRMs introduced in Section \ref{sec:niche}. We encourage the reader to consult \cite{Marsland2019a} for further details and additional examples.

\begin{figure*}
\centering
	\includegraphics[width=0.78\textwidth]{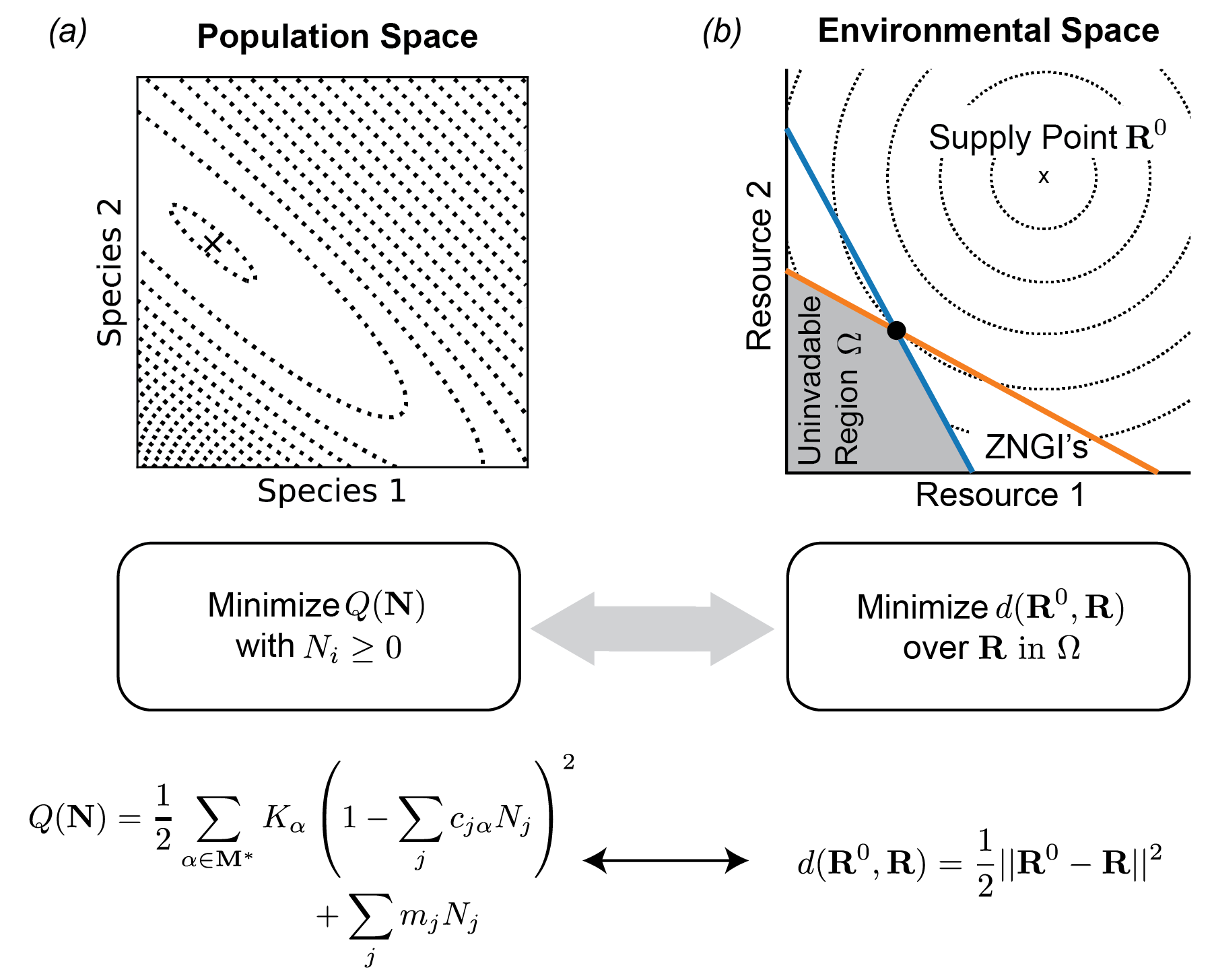}
	\caption{{\bf Reinterpreting MacArthur's Minimization Principle.} \emph{(a)} MacArthur's objective function $Q(\mathbf{N})$ (eqs.~\ref{eq:QN}-\ref{eq:QN1}) plotted as a function of the two population sizes. The `x' shows the steady state reached by numerically simulating the full model of eq.~(\ref{eq:MacN}-\ref{eq:MacR}) using the parameters $r_\alpha = m_i = w_\alpha = 1$ ($\alpha = 1,2$, $i = 1,2$), $K_1 = 4.8$, $K_2 = 2.85$, $c_{1\alpha} = (0.5,0.3)$ and $c_{2\alpha} = (0.4,0.6)$. As MacArthur predicted, the simulation ends at the minimum of $Q$. The panel also gives the simpler form of $Q$ obtained when $r_\alpha = w_\alpha = 1$. \emph{(b)} The same ecosystem viewed in resource space. Contours show the environmental perturbation $d(\mathbf{R}^0,\mathbf{R})$, which measures how far the resource abundances are from the supply point $\mathbf{R}^0$. The colored lines are the zero net-growth isoclines (ZNGIs) of the two species, and the shaded region $\Omega$ is where no species can invade (i.e. $g_i(\mathbf{R})\leq 0$ for all $i$). The steady state (black dot) is the point in $\Omega$ that minimizes $d$. Since $r_\alpha = w_\alpha = 1$, here $d$ is just the Euclidean distance. Figure adapted from Ref.~\cite{Marsland2019a}.}
	\label{fig:intro}
\end{figure*}

\subsection{MacArthur's Minimization Principle}
MacArthur noticed that the GLV representation of the MCRM given above in eq. (\ref{eq:GLV}) and (\ref{eq:riMCRM}-\ref{eq:alphaMCRM}) has a special structure: the per-capita growth rate of each species is proportional to the derivative of a single quadratic function of the population sizes $N_i$:
\begin{align}
\label{eq:QN}
\frac{dN_i}{dt} &= - e_i N_i \frac{\partial Q}{\partial N_i}
\end{align}
with
\begin{align}
Q(\mathbf{N}) &= \frac{1}{2} \sum_{\alpha \in \mathbf{M}^*} r_\alpha^{-1} K_\alpha w_\alpha \left(r_\alpha - \sum_j c_{j\alpha} N_j \right)^2 + \sum_j m_j N_j.
\label{eq:QN1}
\end{align}
To check this, simply take the partial derivative of $Q$ with respect to $N_i$ and compare it to the GLV equations. Notice what eq. (\ref{eq:QN}) tells us about the steady state. Any species that survives ($N_i>0$) must sit at a point where $\partial Q/\partial N_i = 0$. Because of the minus sign in eq. (\ref{eq:QN}), the dynamics always move downhill in $Q$, so this point is a minimum and not a maximum. Species that have gone extinct ($N_i=0$) must not be able to re-invade the community. This requires that $\partial Q/\partial N_i > 0$ (i.e. increasing $N_i$ would increase $Q$). In other words, $Q$ is also minimized along these directions, this time at the boundary $N_i = 0$ set by the requirement that populations be non-negative \cite{gatto1982comments,gatto1990general}. Fig.~\ref{fig:intro}(\emph{a}) shows $Q(\mathbf{N})$ for an ecosystem with two consumers. The `x' marks where a direct numerical simulation of the full MCRM (eq. \ref{eq:MacN}-\ref{eq:MacR}) ends up. As expected, it lies at the minimum of $Q$.

MacArthur's observation tells us a great deal about the steady states of this model. For example, since $Q$ is convex, it has only one local minimum, and hence the model has a unique stable steady state. However, the objective function $Q$ lacks an intuitive interpretation except in some special cases, and the essential assumptions required to obtain an optimization principle remained unclear. In the following sections, we show how the optimization perspective can be applied to a surprisingly large class of models, with a uniform physical interpretation in terms of how far the environment is pushed away from the supply point. The crucial assumption turns out to be the symmetry of the environmentally-mediated interactions among species \cite{Marsland2019a}. In modern language, this is simply the statement that interactions between species are reciprocal.

It is also possible to generalize this optimization framework to models where interactions are not reciprocal. However, in this case, using it to actually find the steady state is more involved (see Section 6.4 below and \cite{Marsland2019a} for a detailed discussion).

\subsection{MacArthur's Consumer Resource Model}
We begin with MacArthur's original model, presented in eqs.~(\ref{eq:MacN}-\ref{eq:MacR}) above. 
At steady-state, we have that
\begin{align}
0 &= N_i  g_i(\mathbf{R})= N_i e_i \left[ \sum_{\beta} w_\beta c_{i\beta} R_\beta - m_i\right] \\
0 &= \frac{r_\alpha}{K_\alpha} R_\alpha(K_\alpha - R_\alpha) - \sum_{j} c_{j\alpha} N_j R_\alpha \label{eq:MacRLag},
\end{align}
where in the first line we have defined the growth rate of species $i$: 
\be
g_i(\mathbf{R})=  e_i \left[ \sum_{\beta} w_\beta c_{i\beta} R_\beta - m_i\right]. 
\ee
To proceed, we rewrite the second of these equations in terms of the function
\begin{align}
d(\mathbf{R}^0,\mathbf{R}) &= \frac{1}{2}\sum_\alpha w_\alpha r_\alpha K_\alpha^{-1} (R_\alpha - R_\alpha^0)^2. \label{eq:dmcrm}
\end{align}
It turns out that $d(\mathbf{R}^0,\mathbf{R})$ is the natural measure of environmental perturbation and can be thought of as a (weighted) Euclidean distance between the resource abundances $\mathbf{R}$ and the supply point $\mathbf{R}^0$. For the MCRM, the supply point is just given by the carrying capacities of the resources, $R_\alpha^0 = K_\alpha$. A straightforward calculation shows that we can rewrite Eq.~\ref{eq:MacRLag} as
\be
0 = \frac{\partial d}{\partial R_\alpha} +\sum_j  \frac{N_j}{e_j}  \frac{\partial g_j}{\partial R_\alpha}.
\ee

If we in addition note that both species and resource abundances must be positive, the steady-state conditions can be summarized by the following four conditions:
\begin{align}
\mathrm{ Steady \,\,populations \,\,} \quad 0 &= \frac{dN_i}{dt} \propto \frac{N_i}{e_i} g_i(\mathbf{R})\label{eq:kkt1}\\
\mathrm{ Steady \,\,environment\,\,}\quad 0 &= -\frac{\partial d}{\partial R_\alpha} - \sum_j  \frac{N_j}{e_j}  \frac{\partial g_j}{\partial R_\alpha}\\
\mathrm{ Noninvasibility\,\,} \quad0 &\geq g_i(\mathbf{R})\\
\mathrm{ Feasible \,\,populations\,\,} \quad 0 &\leq \frac{N_i}{e_i}. \label{eq:kkt4}
\end{align}
Readers familiar with optimization theory will recognize these four equations. They are exactly the Karush-Kuhn-Tucker (KKT) conditions for minimizing the function $d(\mathbf{R}^0,\mathbf{R})$ subject to the inequality constraints $g_i(\mathbf{R}) \leq 0$ \cite{boyd2004convex,bertsekas1999nonlinear}. The KKT conditions are the natural extension of the method of Lagrange multipliers to problems where the constraints are inequalities rather than equalities. In our ecological problem, the role of the Lagrange multipliers (called KKT multipliers in this context) is played by the rescaled species abundances $N_i/e_i$.

 In particular, these four conditions are sufficient to show that the steady-state resource abundances $\mathbf{R}$ are solutions to the following constrained optimization problem
\begin{equation}
\begin{array}{rrclcl}
\displaystyle \min_{R} & \multicolumn{3}{l}{d(\mathbf{R}^0,\mathbf{R})}\\
\textrm{s.t.} & g_j(\mathbf{R}) \le 0\\
&R_\alpha \geq0,   \\
\end{array}
\end{equation}
with $\mathbf{R}^0=\mathbf{K}$.  Since the resource dynamics in eq.~(\ref{eq:MacRLag}) generally has two potential solutions: $R_\alpha=0$ and $R_\alpha>0$, we impose additional constraints $R_\alpha \ge 0$ in the optimization process to find the steady-state solution with the maximum number of surviving resources.

Let us pause to understand the ecological meaning of this objective function. Notice that $d(\mathbf{R}^0,\mathbf{R})$ does not treat all resources equally. Instead, each resource enters with its own weight $w_\alpha r_\alpha K_\alpha^{-1}$, which sets how much a given change in its abundance ``costs''. It is helpful to think about each of the three factors separately. The factor $w_\alpha$ is the nutritional value of resource $\alpha$. If $w_\alpha$ is small, the resource does little to support consumer growth, and it does not matter much if its abundance shifts. The factor $r_\alpha$ is the rate at which the resource regenerates. A resource that grows back quickly is hard to deplete, so a shift of a given size represents a larger disturbance than the same shift in a resource that regenerates slowly. Finally, the factor $K_\alpha^{-1}$ tells us that what matters is the size of a perturbation relative to the carrying capacity. In other words, a change of fixed absolute size is less important for a resource with a large carrying capacity.

In the constrained optimization problem, the species abundances $N_i$ are simply the generalized Lagrange multipliers (KKT multipliers) corresponding to the $S$ inequality constraints $g_i(\mathbf{R}) \le 0$. Notice that the steady-state species condition (known as the K.K.T. condition
in the constrained optimization literature) is simply the statement that $\frac{N_i}{e_i} g_i(\mathbf{R})=0$. It states that species is either extinct (i.e. $g_i(\mathbf{R})<0$ meaning that $N_i=0$) or that a species survives at steady-state (i.e. $g_i(\mathbf{R})=0$ meaning that $N_i>0$). Furthermore, from the definition of Lagrange multipliers, we see that the species abundance $N_i$ can also be thought of as how much the objective $d(\mathbf{R},\mathbf{R}^0)$ changes if we infinitesimally change the growth rate $g_i(\mathbf{R})$ of species $i$.

\subsection{Externally supplied resources}
The environment-based perspective on optimization facilitates generalization to other models not considered by MacArthur, including the model of externally supplied susbstitutable resources discussed above in eq. (\ref{eq:extN}-\ref{eq:extR}). For this model, $d$ is no longer a quadratic function. Instead, it takes the form of a weighted Kullback-Leibler (KL) divergence (see \cite{Marsland2019a} for detailed derivation):
\begin{align}
d(\mathbf{R}^0,\mathbf{R}) &= \omega \sum_\alpha w_\alpha\left[R_\alpha^0 \ln \frac{R_\alpha^0}{R_\alpha} - (R_\alpha^0 - R_\alpha)\right] \label{eq:KL}
\end{align}
where the supply point is given by $R_\alpha^0 = \kappa_\alpha/\omega$. The constrained minimization of this function at equilibrium can easily be verified by taking the derivative and substituting into the KKT conditions as given above in eq. (\ref{eq:kkt1}-\ref{eq:kkt4}). The appearance of the KL divergence here is quite natural. It is commonly used to measure how different two vectors with positive entries are from each other (e.g. two probability distributions or two sets of chemical concentrations) \cite{rao2016nonequilibrium}. Just like in MacArthur's model, the nutritional value $w_\alpha$ sets how much each resource counts. There is, however, one important difference. Since $d(\mathbf{R}^0,\mathbf{R})$ blows up as $R_\alpha \to 0$, the minimum can never lie on the boundary $R_\alpha=0$. In other words, for externally supplied resources we no longer need to impose the constraint $R_\alpha \geq 0$ by hand -- it is automatically satisfied.
	
\subsection{Essential resources}
We now turn to the case of nonsubstitutable resources with fixed biomass stoichiometry, as represented by eq. (\ref{eq:essN}-\ref{eq:essR}) above with $n<-1$. The environmentally mediated interactions among consumers in these models are non-reciprocal because a species can deplete resources that do not limit its own growth. In other words, how a species changes the environment is no longer tied to how the environment affects its growth. In particular, the limit $n\to -\infty$ (``Liebig's Law of the Minimum'') gives the simplest qualitative example of this since each species has a single limiting resource that controls the growth rate and is insensitive to the concentrations of all the other resources. It can be shown that in this limit, the steady-state resource abundances minimize the same objective functions as above (eq. (\ref{eq:KL})), but with the supply point $\mathbf{R}^0$ replaced by an effective supply point $\tilde{\mathbf{R}}^0$.  The effective supply point $\tilde{\mathbf{R}}^0$ corrects the external supply vectors $\mathbf{R}^0$  by accounting for depletion of non-limiting resources at steady state as follows:
\begin{align}
    \tilde{R}_\alpha^0 = R_\alpha^0 - \omega^{-1} \sum_{i,\alpha\neq \beta_i} N_i^* v_{i\alpha} c_{i\alpha}R_\alpha^*\label{eq:Rtild},
\end{align}
where the sum is over all species $i$ that are limited by some resource $\beta_i$ other than $\alpha$. The constrained minimization of $d(\mathbf{R},\tilde{\mathbf{R}}^0)$ in equilibrium can be verified as before by simply substituting into the KKT conditions (\ref{eq:kkt1}-\ref{eq:kkt4}). 

While the notion of the effective supply point defined in eq. (\ref{eq:Rtild}) allows the equilibrium conditions to be formally rewritten as an optimization problem, it also makes the objective function depend explicitly on the solution to the problem. A similar issue is encountered in machine-learning models with latent variables, where the most likely value of the latent variable is itself a function of the best-fit model parameters \cite{mehta2018high}. In machine learning, this conundrum is solved through an iterative algorithm called Expectation Maximization (EM), where the model fitting is performed at each iteration with the current estimate of the most likely latent variable values, and then the latent variables are updated using the new model parameters. An analogous approach proves empirically successful on a range of asymmetric ecological models, including models of essential resources \cite{Marsland2019a}.

%% file: 4_chapter_four.tex
\section{Ecology in high-dimensions}

\subsection{Making the leap to high-dimensions}

Natural ecosystems often have an astounding degree of phylogenetic and physiological diversity. For example,  microbial communities are extremely diverse, ranging from 500-1000 species in human guts \cite{huttenhower2012structure} to over $10^3$ species in marine ecosystems \cite{sunagawa2015structure}. Recent advances in DNA sequencing technologies makes it possible to measure microbial communities abundances at high resolution and has opened a precision era in microbial ecology \cite{caporaso2011global,saliba2014single}. Understanding and modeling such complex datasets challenges current theories and analytical approaches. This has motivated the statistical physics community to start developing a theory of high-dimensional ecosystems.

In low dimensional ecosystems, carefully measuring the traits of each individual in the ecosystems is tractable. However, in high dimensions, this approach becomes intractable for both practical and theoretical reasons. Practically, it is impossible to characterize the preferences and interactions of hundreds of species. Theoretically, even if we could carry out this task, it becomes very difficult to make senes of the resulting high-dimensional dynamics.

A similar problem is encountered in physics when modeling systems such as gases which are composed of a large ensemble of particles. One pertinent example from statistical mechanics is the example of  \textsl{particles in a box}.  At any time $t$, the microstate of a system of $N$ particles is described by $6N$ numbers, namely  the positions $\vec{x}_i(t)$  and the momenta $\vec{p}_i(t)$, of the $i=1,\ldots N$ particles. When $N$ is small, the evolution of the microstate can be predicted precisely by solving Hamilton's equations for all particles. However, when $N \gg 1$ ( e.g. for $N \sim 10^{23}$ as is the case for a gas),  the complexity of predicting the microstate is too high to be feasible. Instead, one is forced to describe the system \emph{statistically}.  Making an analogy between particles in a box and species in an ecosystem,  this suggests that analytical approaches and theoretical insights derived from small ecosystems with a few species and resources may not scale up to large, diverse ecosystems. 

We have stressed the practical difficulties in high dimension. Actually, taking the thermodynamic limit $N\rightarrow \infty$ can also lead to a number of simplifications. As we all know, statistical mechanics provides a prescription for successfully dealing with systems with a very large number of degrees of freedom \cite{ma2018modern}. For the \textsl{particles in a box} example, instead of trying to predict the behavior of individual particles, we can make accurate statements about the macroscopic quantities such as temperature and pressure which reflect averages over millions of particles. As a result of this averaging, macroscopic quantities often have universal behaviors. For example, the pressure, temperature, volume and entropy are related by Maxwell's relations regardless of if the gas particles being analyzed are  oxygen, nitrogen or mixture of the two. This reflects the fact that in the thermodynamics limit, the behaviors we will observe will be \emph{typical}.

The analogy between species in ecosystems and particles in box yields some general lessons from statistical mechanics that we can use to  study high-dimensional ecology. First, we should concentrate of finding relationship between macroscopic properties of the ecosystem rather than the behaviors of any particular species or resource. Second, we should focus on characterizing the typical behaviors we would expect to see in diverse ecosystems. Finally, on a technical level, we must exploit the central limit theorem by averaging over species and resource traits to make predictions about ensembles of ecosystems rather than any particular ecosystem.

\subsection{May's Stability Criteria and Random Matrix Theory}\label{sec:May}

\begin{figure}[h]
\centering
\includegraphics[width=0.58\textwidth]{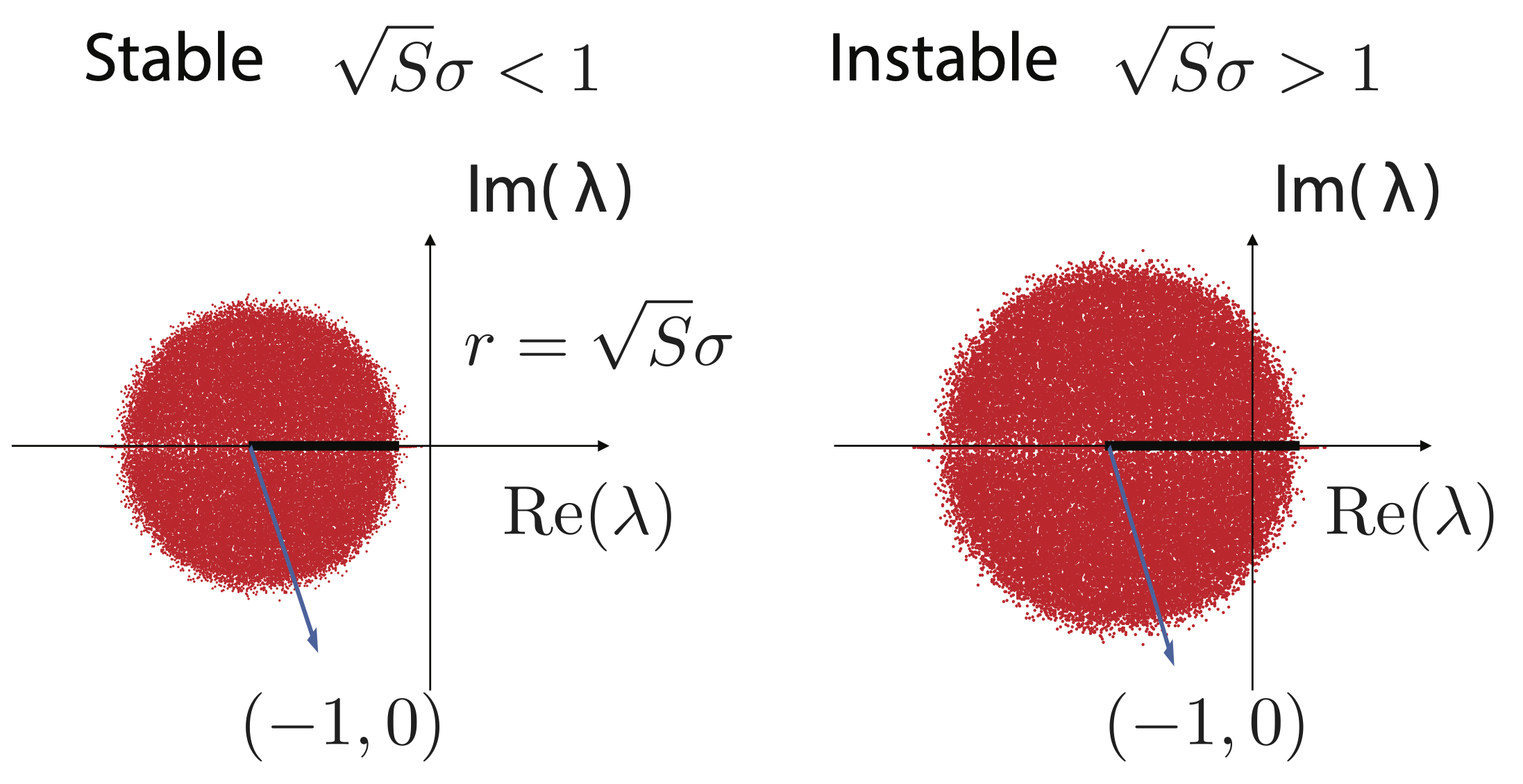}
\caption{Schematic for May's stability criteria. The red scatter points are the eigenvalues on the complex plane. }
\label{Fig:May}
\end{figure}

One of the first examples of the use of ideas from statistical mechanics to ecology was the pioneering work by physicist-turned-ecologist Robert May  \cite{may1972will}. May was inspired by Wigner's work on Uranium showing that one could recapitulate many aspects of experimental data by replacing the exact Uranium Hamiltonian by a  large symmetric random matrix. The underlying reason for this was that the predicted properties of Uranium were \emph{typical}. In other words, these properties did not depend on the detailed interactions of Uranium but instead were properties of any sufficiently complex system. Inspired by this, Wigner showed that one could model some aspects of complex Hamiltonians (i.e. the level statics) with large random matrices.

May transplanted these ideas to an ecological setting. He considered a general ecosystem with $S$ species, with abundances $N_i$, whose dynamics take the form
\begin{equation}\label{eq:Maygen}
    \frac{d N_i}{dt}= g_i(N_1, N_2, \ldots, N_S)=g_i(\mathbf{N})
\end{equation}
He further assumed that the ecosystem had a fixed point given by $\bar{\mathbf{N}}$. He then asked about the nature of the dynamics around the fixed point.  In this setting, the population dynamics can be described by Taylor expanding the dynamics around the equilibrium point $\bar{\mathbf{N}}$ to get
\begin{equation}\label{eq:May}
    \frac{d N_i}{dt}=\sum_j J_{ij}(N_j -\bar{N}_j).
\end{equation}
May wanted to know whether the fixed point $\bar{\mathbf{N}}$ is stable when the number of species is large ($S \gg 1$). Since eq.~(\ref{eq:May}) is linear, the answer is determined by the eigenvalue $\lambda_{\rm max}$ of $J_{ij}$ with the largest real part. If $\lambda_{\rm max}$ is negative, any small perturbation decays and the system returns to $\bar{\mathbf{N}}$. In contrast, if $\lambda_{\rm max}$ is positive, small perturbations grow and the ecosystem is driven away from the fixed point (i.e. the fixed point is unstable).

To answer this question, May made use of results from Random Matrix Theory. In the 1960's, Jean Ginibre calculated the eigenvalue distribution of large random matrices whose entries are independent Gaussian random variables \cite{ginibre1965statistical}. This result was later generalized by Girko and is now known as the Circular Law. It says that if the entries $J_{ij}$ are drawn independently from a distribution with zero mean and variance $\sigma^2$, then as $S\rightarrow \infty$ the eigenvalues uniformly fill a disk of radius $r=\sqrt{S}\sigma$ in the complex plane (see Fig. \ref{Fig:May}). In particular, the eigenvalue with the largest real part sits on the edge of this disk.

May then performed a simple thought experiment. He started with an ecosystem of $S$ species that do not interact with each other, but where each species limits its own growth (i.e. $J_{ii} = -1$ and $J_{ij}=0$ for $i \neq j$). All the eigenvalues of this matrix equal $-1$, so this ecosystem is stable no matter how many species it contains. He then asked what happens when we turn on interactions between species by drawing the off-diagonal elements $J_{ij}$ at random. The $-1$'s on the diagonal simply shift the disk of eigenvalues to the left by one. Using the Circular Law, we then find that the eigenvalue with the largest real part is
\begin{align}
\lambda_{\rm max} &\sim -1 +  \sqrt{S \sigma^2}.
\end{align}
 In other words, the ecosystem becomes unstable once the typical combined effect of all the other species on a given species, $\sqrt{S\sigma^2}$, is as large as the self-regulation of that species. This leads to May's famous stability criterion for the largest number of species $S^*$ that can stably coexist:
\begin{equation}\label{eq:Maycriteria}
 \sqrt{S^* \sigma^2} = 1.
\end{equation}
Thus, according to May, the more strongly species interact (i.e. the larger $\sigma$), the fewer species an ecosystem can support before it becomes unstable.

As is clear from the argument above, a large number of assumptions were made in this derivation. Nonetheless, this result proved to be extremely influential.  Before the 1970s, ecologists believed that diversity enhanced ecosystem stability. May's stability criteria challenged this idea and lead to what is now known as the ``diversity-stability debate'' \cite{mccann2000diversity}. Since May's original publications, theorists showed how this criteria could be violated  by changing the May's original assumptions, including but not limited to, adding
biologically realistic correlation structures \cite{allesina2012stability}, modular structures \cite{grilli2016modularity},  accounting for the fact that the community
matrix depends on population sizes \cite{gibbs2018effect}, and considering high-order interactions Lotka–Volterra dynamics \cite{grilli2017higher}.

\section{Cavity Method for Lotka–Volterra model}\label{sec:LV}
\begin{figure}
\centering
\includegraphics[width=0.70\textwidth]{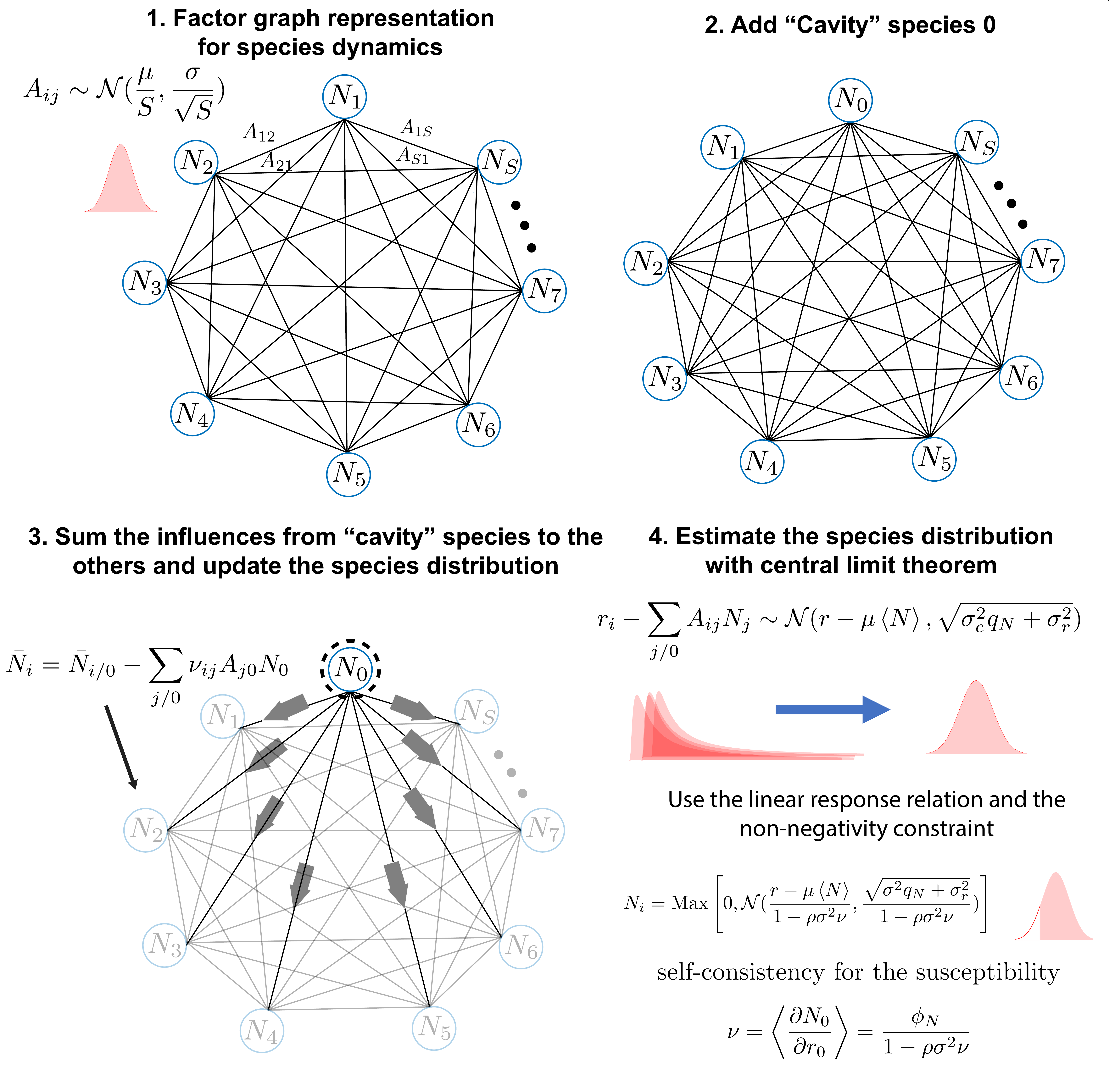}
\caption{Schematic outlining steps in cavity solution for Lotka–Volterra model. \textbf{1.} The species dynamics in eq. (\ref{eq:LVeq}) are expressed as a factor graph. The edges are bi-directional and sampled from a Gaussian distribution.   \textbf{2.} Add  the "Cavity" species $0$ as the perturbation. \textbf{3.} Sum the resource abundance perturbations from the "Cavity" species 0 at steady state and update the species abundance distribution to reflect the new steady state. \textbf{4.}  Employing the central limit theorem and the non-negativity constraint, the species distribution is expressed as a truncated normal distribution.  The susceptibility appearing in the species distribution is the self-consistency relation. }
\label{cavity:schemeLV}
\end{figure}

One major shortcoming of May's argument is that he considered purely linearized dynamics. Over the last few years, physicists have extended these ideas to more complex ecological models such including the Generalized  Lotka-Volterra model (GLV) and the consumer resource models (CRMs). One crucial difference between May's linear dynamics and these latter two classes of models is that in both GLVs and CRMs, species can go extinct. This gives rise to additional self-organization due to species extinctions that fundamentally change the qualitative picture derived by May.

We begin by outlining the cavity solution to the GLV. Our derivation closely follows \cite{bunin2017ecological}, though we deploy what we consider to be more transparent notation. We consider a GLV described by $S$ equations of the form
\begin{equation}\label{eq:LVcavity}
\frac{dN_i}{dt}=g_i N_i(r_i -N_i -\sum_j A_{ij}N_j), \quad i=1,2, \dots, S.
\end{equation}
Here $g_i$ are the intrinsic growth rates, $r_i$ are the carrying capacities and $A_{ij}$ encode inter-species interactions, with positive and negative values representing competition and mutualism interactions, respectively. We are interested in characterizing the steady-state  solutions to these equations (i.e.  $\frac{dN_i}{dt}=0$)  when the  $A_{ij}$ are sampled from a random distribution. In other words we would like to characterize the statistical properties of $\{ {N}_j\}$ that satisfy the fixed point equations
\begin{equation}\label{eq:LVeq}
0= {N}_i(r_i -{N}_i -\sum_{j\neq i} A_{ij}{N}_j).
\end{equation}

To do so, we will make use of the zero temperature cavity method \cite{mezard2003cavity}. The cavity method has been successfully used in a wide variety of settings, ranging from spin glasses  to computer science  and combinatorial optimization. Over the last five or six years, there has been a concerted effort to develop and generalize the cavity method to solve ecological models with non-negative continuous variables (species and resource abundance) \cite{bunin2017ecological, Advani2018, barbier2017cavity, mehta2021cross}. 

The basic idea behind the cavity method is to derive  self-consistency equations by relating an ecosystem with S species to another ecosystem with  S+1 species where the interaction coefficients $A_{ij}$ are drawn from the same random distribution (see Figure ~\ref{cavity:schemeLV}). In the thermodynamic limit $S \rightarrow \infty$, the macroscopic observables of these two systems will be indistinguishable. In other words, statistical observables such as the first and second moment of species abundances $\left<N \right>$, $\left<N^2 \right>$ in the two ecosystems, will be identical up to order $\sim 1/S$ corrections which become negligible in the thermodynamic limit. 

Technically, we do this by adding a new ``cavity" species 0  to the original ecosystem. The addition of this species can be viewed as a small perturbation to the original equilibrium state and hence can be calculated using perturbation theory. This allows us to derive self-consistency equations characterizing the observables of interest. In what follows, we assume that the $A_{ij}$ are sampled from a Gaussian distribution with mean $\left< A_{ij}\right>= \frac{\mu}{S}$ and variance $var(A_{ij})= \frac{\sigma^2}{S}$. The Gaussian assumption is not essential. In fact, what is important is that the distribution is not long-tailed with finite first two moments give by the expression above.

In order to employ the cavity method, it is helpful to rewrite
 \begin{equation}\label{eq:A}
     A_{ij}=\frac{\mu}{S}+\sigma a_{ij}
 \end{equation}
 with 
  \begin{equation}
 \begin{aligned}
 \left<a_{ij} \right>&=0 \\
 \left<a_{ij}a_{kl} \right>&=\frac{1}{S}\delta_{ik}\delta_{jl} +\frac{\rho}{S}\delta_{il}\delta_{jk},
 \label{eq:aijstats}
 \end{aligned}
 \end{equation}
 where $ -1 \le \rho \le 1$ measures the amount of non-reciprocity between species-species interactions. In particular, $\rho=1$ corresponds to the case of reciprocal interactions where species $i$ affects species $j$ the same way species $j$ affects species $i$. The scaling in the mean and variance is chosen to keep the interaction term $\sum_j A_{ij}\bar{N}_j$ in eq. (\ref{eq:LVeq}) independent of system size. 
 
 The parameters $r_i$, the carrying capacities in the GLV,  are also sampled from a Gaussian distribution with mean $r$ and variance $\sigma_r$. Importantly, we assume that the $r_i$ and $A_{ij}$ are uncorrelated. To aid calculations, it is helpful to write
 \begin{equation}\label{eq:r}
    r_i=r+\delta r_i,
\end{equation}
 where 
\begin{equation}
 \begin{aligned}
 \left<\delta r_i \right>&=0  \\ 
 \left<\delta r_i \delta r_j \right>&=\sigma_r^2 \delta{ij}.
 \label{eq:ristats}
 \end{aligned}
\end{equation}


\subsection{The Cavity Solution}

To perform the cavity calculation, we perturb the original ecosystem with a new ``cavity" species $N_0$ with interaction coefficient $A_{0j}$ and $A_{j0}$ (see  Figure ~\ref{cavity:schemeLV}). The equations for this new ecosystem with $S+1$ species take the form 
\begin{eqnarray}\label{N0_wc}
\begin{aligned}
0 &= {N}_i(r_i -{N}_i -\sum_{j\neq i} A_{ij}{N}_j -A_{i0}N_0) \\
0 &= N_0(r_0-N_0- \sum_{j=1}^SA_{0j}N_{j}),
\end{aligned}
\end{eqnarray} 
where for notational simplicity we have dropped the bar indicating steady-state. 

The key observation is that since $A_{i0}$ is of order $1/S$, the term $A_{i0}N_0$ is a small perturbation whose effects can be calculated perturbatively. In particular, from the viewpoint of a species $i$, the addition of species zero is the same as slightly changing $r_i$, namely $r_i \rightarrow r_i  + \delta r_i$ with $\delta r_i =-A_{i0}N_0$. Motivated by this observation, we define 
the susceptibility function
\be
\nu_{ij}=\frac{\partial N_i}{\partial r_j}.
\label{eq:defmatrixsus}
\ee
If we denote the steady state abundances in the original $S$-species ecosystem by ${N}_{i/0}$ and the steady state abundances in the $(S+1)$-species ecosystem after the addition of species $0$ by $N_{i}$,  first order perturbation theory tell us that these two quantities are related by the expression 
\begin{eqnarray}
N_i &=& {N}_{i /0} -\sum_{j=1}^S\nu_{ij} A_{j0} {N}_0.
\label{eq:perturbNi}
\end{eqnarray}
We can now substitute this relationship into the second equation in eq. (\ref{N0_wc}) to get
\begin{eqnarray}
0= N_0(r_0-N_0- \sum_{j=1}^SA_{0j} (N_{j /0} -\sum_{i=1}^S\nu_{ji} A_{i0} {N}_0)).
\label{eq:N0eqcavity}
\end{eqnarray} 
Substituting eq. (\ref{eq:A}) and eq. (\ref{eq:r}) and solving for $N_0$ gives two potential solutions over the reals: $N_0=0$ and 
\begin{equation}
N_0=\frac{r +\delta r_0-\sum_{j=1}^S\sigma a_{0j}{N}_{j /0}-\mu\left <N \right> }{1- \sigma^2\sum_{j,i=1}^S\nu_{ij} a_{0j}a_{i0} },
\end{equation}
where we have defined $\left< N\right>=\frac{1}{S}\sum_j N_j$ and we have ignored all terms of order $1/S$. Since species abundances must be \emph{positive}, the latter solution may not exist if the resulting expression for $N_0$ is negative. If we also require that the solution be noninvasible (i.e. that if a positive solution exist we choose that solution), we can solve for $N_0$ to get
\begin{eqnarray}\label{eq:LVsolN1}
 N_0=\mathrm{max}\left(0, \frac{r +\delta r_0-\sum_{j=1}^S\sigma a_{0j}{N}_{j /0}-\mu\left <N \right> }{1- \sigma^2\sum_{j,i=1}^S\nu_{ij} a_{0j}a_{i0} }\right)
\end{eqnarray}
The $\mathrm{max}$ function simply takes the maximum value in the bracket. 
To leading order in $1/S$, we can replace the  $a_{0j}a_{i0}$ in the double sum by its expectation values so that
\begin{equation}
\sum_{j,i=1}^S\nu_{ij} a_{0j}a_{i0}=\frac{\rho}{S}\sum_{j}\nu_{jj}=\rho\nu,
\end{equation}
where we have defined the average trace of the susceptibility matrix
\be
\nu ={1 \over S}\sum_{j}\nu_{jj}.
\label{Eq:deftracenu}
\ee
Substituting this into Eq.~\ref{eq:LVsolN1} , we get
\begin{eqnarray}\label{eq:LVsolN}
 N_0=\mathrm{max}\left(0, \frac{r +\delta r_0-\sum_{j=1}^S\sigma a_{0j}{N}_{j /0}-\mu\left <N \right> }{1- \sigma^2 \rho \nu }\right).
\end{eqnarray}

So far, we have not done anything out of the ordinary. However, at this point we will make use of the power of ``self-averaging'' and typicality. Instead, of characterizing the solutions to Eq.~\ref{eq:LVsolN1} for a particular choice of $\delta r_i$ and $a_{ij}$, we will instead ask about the property of solutions ``averaged'' over the different realizations of these variables. In other words, we imagine solving Eq.~\ref{eq:LVsolN1} for many different choices of $a_{ij}$ and $r_i$, where these variables are all drawn from the same distribution, and ask about the distribution of solutions of $N_0$ for these different choices. Rather than focus on the full distribution, we will characterize the distribution by three quantities: the fraction of times that $N_0$ is non-zero, $\phi_{N_0}$,  the mean value of the species abundance $\<N_0\>$, and the second moment of  the species distribution $\<N_0^2\>$. Here, the brackets denote averages over different realizations of the random interactions $a_{ij}$ and $\delta r_i$.

Since we are interested in the behavior of solutions over different realizations of the random interactions, we can use the central limit theorem to further simplify Eq.~\ref{eq:LVsolN1}.  It is clear that if we make a plot of the the sum  $\sum_{j=1}^S\sigma a_{0j}{N}_{j /0}$ in the numerator over different realizations of the $a_{0j}$, it will be have like a Gaussian random variable with mean
\be
\<\sum_{j=1}^S\sigma a_{0j}{N}_{j /0}\>= S\sigma \<a_{0j}\> \<N\>=0,
\ee
and variance
\be
\<(\sum_{j=1}^S\sigma a_{0j}{N}_{i /0})^2\>=\sigma^2 \sum_{jk} \<a_{0j}a_{0k}\> N_{j/0} N_{k/0}
=\sigma^2 \<N^2\>
\ee
where we have made use of Eq.~\ref{eq:aijstats} and defined the quantities
\begin{equation}
 \begin{aligned}
\<N\> &={1 \over S} \sum_{i}N_{i/0}  \\
\<N^2\> &={1 \over S} \sum_{i}N_{i/0}^2.
\label{eq:defNN2}
 \end{aligned}
\end{equation}
Furthermore, by assumption $\delta r_i$ is a random normal variable with variance $\sigma_r^2$ that is uncorrelated with $a_{ij}$. Hence, the quantity $\sum_{j=1}^S\sigma a_{0j}{N}_{i /0} +\delta r_i$ is also a random normal variable with mean $0$ and variance given by the sum of the variances of the individual terms: $\sigma_r^2 + \sigma^2 \<N^2\>$. 

We can combine these observations with eq. (\ref{eq:LVsolN}) to write an expression for the distribution of $N_0$ over different realizations of $a_{ij}$ and $\delta r_i$. Namely, if we define $z_N$ to be a standard normal random variable with mean $0$ and standard deviation $1$, then the distribution of solutions for $N_0$ takes the form
\begin{eqnarray}\label{eq:LVsolN2}
 N_0=\mathrm{max}\left(0, \frac{r-\mu\left <N \right> +\sqrt{\sigma_r^2+\sigma^2 \left <N^2 \right> }z_N}{1- \rho \sigma^2 \nu}\right).
\end{eqnarray}
Notice that the the right hand argument in the max 
\be
\frac{r-\mu\left <N \right> +\sqrt{\sigma_r^2+\sigma^2 \left <N^2 \right> }z_N}{1- \rho \sigma^2 \nu}
\ee
is just the formula for a Gaussian distribution with mean $(r-\mu\left <N \right>) /({1- \rho \sigma^2 \nu})$ and variance $(\sigma_r^2+\sigma^2 \left <N^2 \right> )/{(1- \rho \sigma^2 \nu)^2}$. Thus,  the full distribution of $N_0$ given in Eq.~\ref{eq:LVsolN2} is a truncated Gaussian. A truncated Gaussian can be completely characterized by three quantities: the probability that the new species survives $\phi_{N_0}$ (i.e. the probability that $N_0>0$) and its first two moments, $\<N_0\>$ and $\<N_0^2\>$.

Note that Eq.~\ref{eq:LVsolN2} still depends on three unknown quantities:  the average trace of the susceptibility $\nu$, as well as the first and second moments of the species distributions $\<N\>$ and $\<N^2\>$. We emphasize that $\<N\>$ and $\<N^2\>$ are moments over all species in an ecosystem for a \emph{single} realization of the random parameters $a_{ij}$ and $r_i$. The same holds true for $\nu$. In order to make progress, we now invoke the idea of self-averaging. Namely, the statement that as the number of species $S$ goes to infinity, the statistical distribution characterizing $N_i$ for single realization of the random parameters should be identical to the distribution characterizing $N_0$ over many different realizations of the random parameters. In the language of the statistical physics of disordered systems, we focus on the replica symmetric solution. 

Self-averaging and replica symmetry follow from the assumptions that: (i)  the new species we have introduced is statistically indistinguishable from the original $S$ species and (ii) as $S$ gets large, we can think of the entire ecosystem as being composed of smaller ecosystems, each with a different independent realization of the random parameters.  As we will briefly discuss below, the replica symmetric solution is not always valid and more complicated things can happen, especially as one increases the variance of the underlying distribution from which the parameters are drawn. However, the replica-symmetric solution always serves as a good starting point for understanding the behavior of high-dimensional ecosystems.

Assuming replica symmetry and self-averaging, we can equate expectations of $N_0$ computed using the truncated Gaussian distribution in Eq.~\ref{eq:LVsolN2}, with averages calculated over all species from a single realization. This allows us to derive self-consistency equations of the form:
\begin{equation}
 \begin{aligned}
\phi_{N_0}= \phi_N = {1 \over S} \sum_{i}^S \theta(N_{i/0}) \\
\<N_0 \> = \<N\> = {1 \over S} \sum_{i}^S N_{i/0}  \\
 \<N_0^2\>= \<N\> = {1 \over S} \sum_{i}^S N_{i/0}^2   \\
 \left< \frac{\partial N_0}{\partial r_0}\right>= {1 \over S} \sum_{i} \nu_{ii},
  \end{aligned}
\end{equation}
 where in the last line we have used the definition of susceptibility Eqs.~\ref{eq:defmatrixsus} and \ref{Eq:deftracenu}.
Explicitly differentiating Eq.~\ref{eq:LVsolN2} with respect to $r_0$, this las self-consistency relation yields
\begin{equation}
\nu=\left< \frac{\partial N_0}{\partial r_0}\right>=\frac{\phi_N}{1-\rho\sigma^2\nu}.
\end{equation}
where $\phi_N=\frac{S^*}{S}$ is the fraction of nonzero species in the ecosystem. Note that $\phi_N$ results from the fact that if $N_0=0$, the corresponding derivative is zero.

In order to write the first three self-consistent equations more explicitly, it is helpful to define the following notation. Let $y=\text{max}\left(0, \frac{c}{b}z+\frac{a}{b}\right)$, with $z$ being a Gaussian random variable with zero mean and unit variance. Then its $j$-th moment is given by
\begin{equation}
 \begin{aligned}
 \label{eq:wj}
\left<y^j \right>&=&\frac{1}{\sqrt{2\pi}}\int_{-\frac{a}{c}}^\infty e^{-\frac{x^2}{2}}\left (\frac{c}{b}x+\frac{a}{b} \right)^j dx\\
&=&\left(\frac{c}{b}\right)^j\frac{1}{\sqrt{2\pi}}\int_{-\frac{a}{c}}^\infty e^{-\frac{x^2}{2}}\left (x+\frac{a}{c} \right)^j dx\\
&=&\left(\frac{c}{b}\right)^jw_j(\frac{a}{c}),
  \end{aligned}
\end{equation}
where in the last line we have defined the functions 
\be
w_j(\frac{a}{c})=\frac{1}{\sqrt{2\pi}}\int_{-\frac{a}{c}}^\infty e^{-\frac{z^2}{2}}\left (z+\frac{a}{c} \right)^j dz.
\ee
In terms of the $w_j$, the self-consistency equations take the form
\begin{eqnarray}
&&\nu=\frac{\phi_N}{1-\rho\sigma^2\nu},\quad
\phi_N=w_0(\frac{r-\mu\left <N \right> }{\sqrt{\sigma_r^2+\sigma^2 \left <N^2 \right>}}),\label{eq:LVcavity1}\\
&&\left <N \right>=\frac{\sqrt{\sigma_r^2+\sigma^2 \left <N^2 \right>} }{1- \rho \sigma^2 \nu}w_1(\frac{r-\mu\left <N \right> }{\sqrt{\sigma_r^2+\sigma^2 \left <N^2 \right>}}),\label{eq:LVcavity2}\\
&&\left <N^2 \right>=\frac{\sigma_r^2+\sigma^2 \left <N^2 \right> }{(1- \rho \sigma^2 \nu)^2}w_2(\frac{r-\mu\left <N \right> }{\sqrt{\sigma_r^2+\sigma^2 \left <N^2 \right>}})\label{eq:LVcavity3}.
\end{eqnarray}

\begin{figure}
\centering
\includegraphics[width=0.5\textwidth]{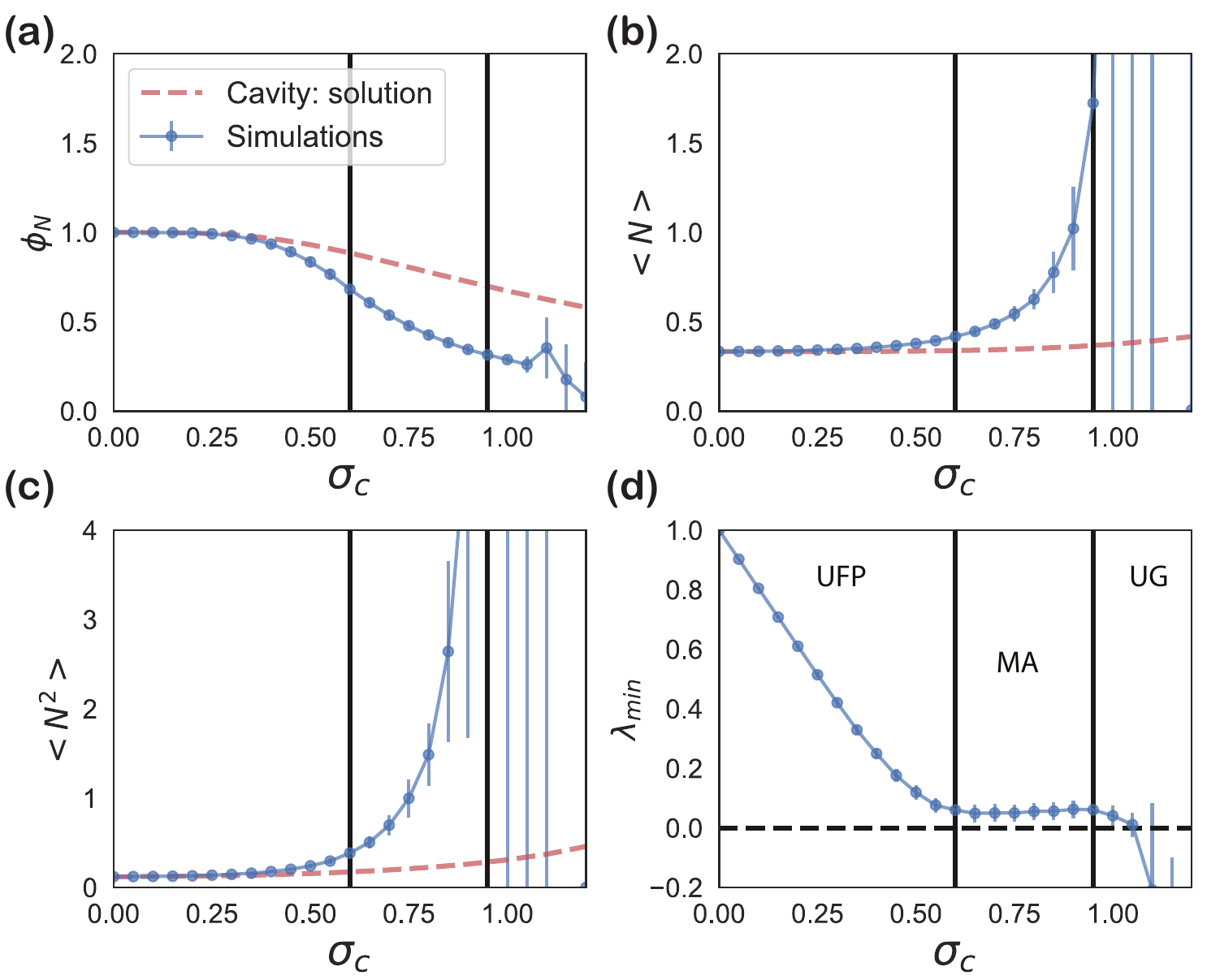}
\caption{Comparison between the cavity solution (equation \ref{eq:LVcavity1} - \ref{eq:LVcavity3}) and  simulations for symmetric interactions ($\rho=1$), (a) the fraction of surviving species $\phi_N= \frac{S^*}{S}$
and (b) the first moment $\left<N\right>$ and (c) the second moment $\left<N^2\right>$ of the species distributions as a function of $\sigma_c$. (d) The minimum eigenvalue of the submatrix $A_{ij}^*$ at at different $\sigma_c$. The error bar shows the standard deviation from $500$ numerical simulations with $S=200$, $\mu=2.$, $r=1.$, $\sigma_r=0.1$ and $\rho=1$.The black solid lines separate the results in three different regimes: unique fixed point, multiple attractors and unbound growth.The black dashed line in Panel (d) is the base line $\lambda=0$ for comparison.  }
\label{fig:cavityLVsolution}
\end{figure}

\begin{figure}[h]
\centering
\includegraphics[width=0.48\textwidth]{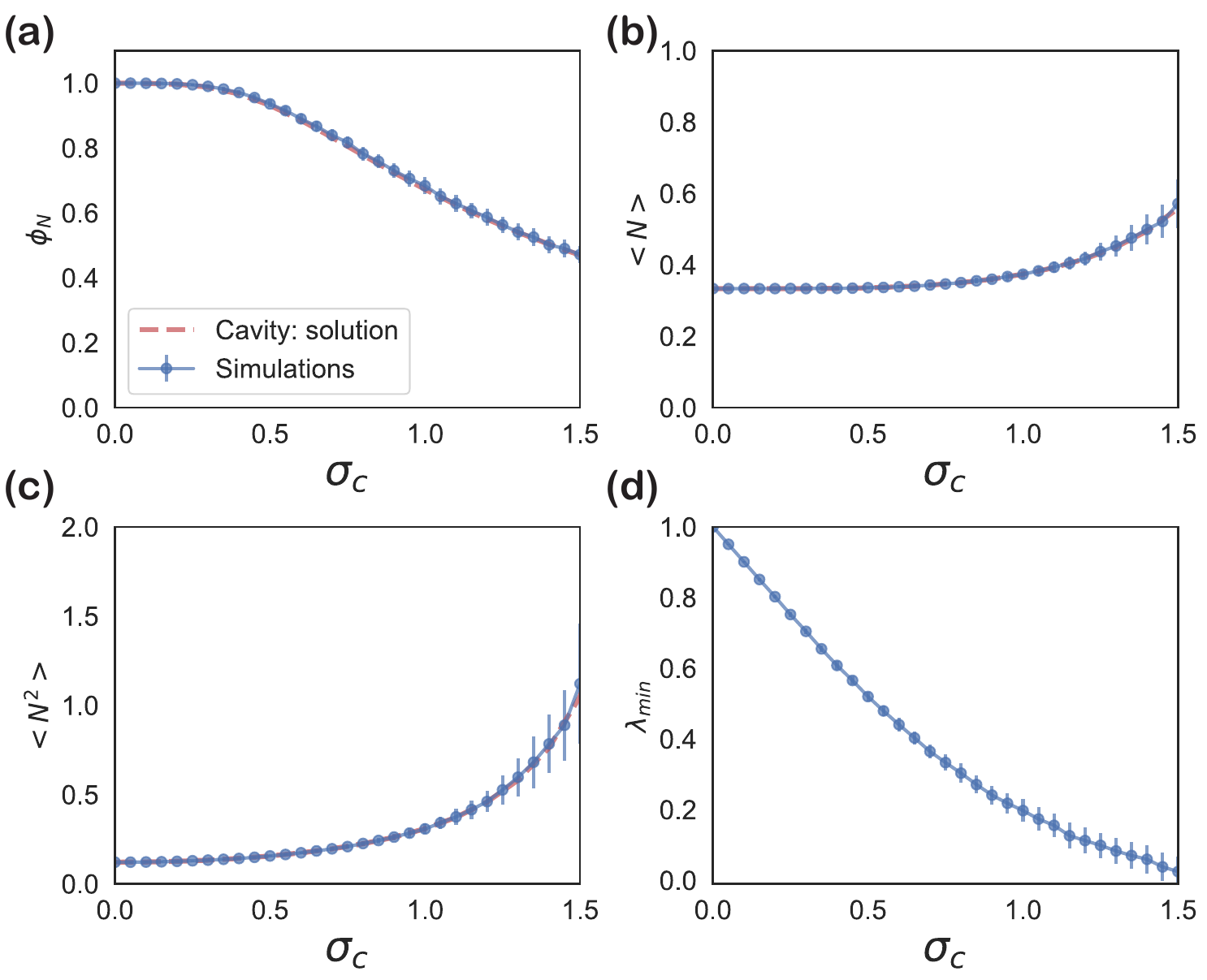}
\caption{Comparison between the cavity solution (equation \ref{eq:LVcavity1} - \ref{eq:LVcavity3}) and  simulations for uncorrelated interactions, $\rho=0$, (a) the fraction of surviving species $\phi_N= \frac{S^*}{S}$
and (b) the first moment $\left<N\right>$ and (c) the second moment $\left<N^2\right>$ of the species distributions as a function of $\sigma_c$. (d) The minimum eigenvalue of the submatrix ${A_{ij}^*}$ at at different $\sigma_c$. The error bar shows the standard deviation from $500$ numerical simulations with $S=200$, $\mu=2.$, $r=1.$, $\sigma_r=0.1$ and $\rho=0$.}
\label{cavity:LVsolution_unsym}
\end{figure}

These self-consistency equations can be
solved numerically in Mathematica. Fig. \ref{fig:cavityLVsolution} and Fig. \ref{cavity:LVsolution_unsym} show the comparison between the cavity solution and 500 independent numerical simulations for various ecosystem properties for symmetric ($\rho=1$) and uncorrelated ($\rho=0$) interactions $A_{ij}$. As can
be seen in the figures, our analytic expressions agree remarkably well over a large range of $\sigma_c$. Notice, for the ecosystem with symmetric interactions (Fig. ~\ref{fig:cavityLVsolution}), the numerics and analytic expression start to disagree for $\sigma_c \sim 0.5$.

\subsection{Beyond Replica Symmetry}

The underlying reason for the disagreement between the analytic predictions and the numerical simulations for ecosystems with symmetric interactions is that the replica symmetric solution becomes unstable for $\sigma_c \sim 0.5$, indicating a phase transition to the ``replica-symmetry broken'' phase. The breaking of replica symmetry is actually closely related to May's original arguments about stability of ecosystems. 

To see this, it is helpful to recall that one important criteria for thinking about whether a system is stable is to ask how sensitively it responds to perturbations in external parameters.  We emphasize that this criteria is closely related to, but different from, asking about the stability of ecosystems to perturbing the species abundances. It is conceptually helpful to consider this distinction in greater mathematical detail. To do so,  we will focus only on the $S^*$ surviving species, rather than all $S$ original species in the regional species pool.  From Eq.~\ref{eq:LVeq}, we know at steady-state we must have (where $i$ runs over the $S^*$ surviving species)
\be 
0={dN_i \over dt} = N_i (r_i - \sum_{k=1}^{S^*}A^*_{ik} N_k),
\label{eq:ssLVsurv}
\ee
where $A^*_{ij}$ denotes the $S^* \times S^*$ interaction sub-matrix of the full interaction matrix $A_{ij}$ restricted to surviving species and the index $i$ runs over the $S^*$ surviving species.

Then, the stability to perturbations in species abundances (the kind of stability considered by May), can be calculated by linearizing this equation to get an equation for the deviations $\delta N_i$ from the steady-state abundances $N_i$:
\be
{d \delta N_i \over dt} = -\sum_{k=1}^{S^*} N_i A^*_{ik} \delta N_k,
\ee
where $\delta_{ik}$ is the Kronecker Delta function (i.e. $S^* \times S^*$ identity matrix). This is the analogue of Eq.~\ref{eq:May}. Notice that this stability is set by the eigenvalues of the matrix $-N_i A^*_{ij}$ which unlike the case May considered depends on the steady-state abundances at the steady-state.

For this reason, it will be helpful to work with a slightly different quantity which ask about the sensitivity of steady-state abundances $N_i$ to small changes in the external parameters $r_j$. This is precisely the susceptibility, Eq.~\ref{eq:defmatrixsus}, but now restricted to surviving species. We denote this restricted susceptibility by $\nu_{ij}^*$. We can solve for it by  dividing Eq.~\ref{eq:ssLVsurv} by $N_i$ and then differentiating by $r_j$ to get
\be
0= \delta_{ij}^* -  \sum_{j=1}^{S^*}A^*_{ik} {dN_k \over dr_j},
\ee
where the Kronecker-delta function $\delta_{ij}^*$ comes from the fact that  ${\partial r_i \over \partial r_j}=\delta_{ij}$. Inverting this equation gives
\be
\nu_{ij}^* = {dN_i \over d r_j}=(A_{ij}^*)^{-1}
\ee
It turns out that RS breaking corresponds to the case where $A_{ij}^*$
develops a zero eigenvalue, indicating that the susceptibility diverges. In other words, the system becomes infinitely sensitive to external perturbations. 

Fig. \ref{fig:cavityLVsolution} (D) shows the minimum eigenvalue of $A_{ij}^*$ as $\sigma_c$ is varied. The minimum eigenvalue decrease monotonically with increasing $\sigma_c$ until it is close to zero. And then the cavity solution fails. This correspond to the emergence of multiple attractors (MA phase). Further increasing $\sigma_c$ results in unbounded growth. The full phase diagram and consequences of this observations were first worked out in \cite{bunin2017ecological}. To analyze these phases, one can recast the problem using the replica approach with replica symmetry breaking \cite{Biroli2017}.

One of the most striking things about the GLV model is that the minimum eigenvalue of $A_{ij}^*$ gets pinned exactly to zero over an extended region of the parameter space (region indicated by MA in Fig. \ref{fig:cavityLVsolution} (D)). This corresponds to the phenomena of marginal stability. In other words, though the system is not unstable, it is infinitely sensitive to small changes in external parameters. This marginal stability is unique to the symmetric GLV model and is an indication that the ecosystem self-organizes through species extinction. Generically with non-reciprocal interactions, the marginal stable phase disappears and is replaced by a dynamic phase that can exhibit chaotic fluctuations 
 \cite{pearce2020stabilization,roy2021glassyLV,roy2020persistent,  ros2023rGLVtypical,ros2023rGLVquenched,pirey2023aging}.

\section{Cavity Method for MacArthur Consumer-Resource Models}\label{sec:GCRM}

\begin{figure*}
\centering
\includegraphics[width=0.88\textwidth]{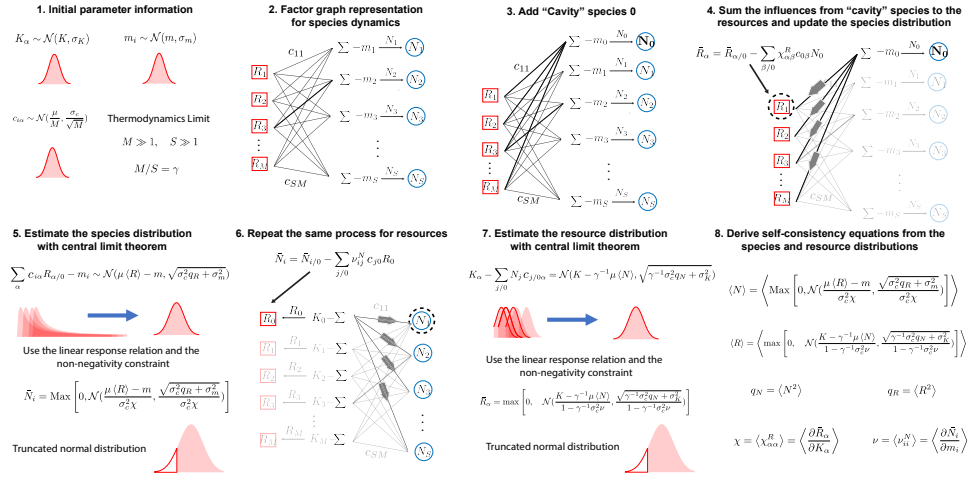}
\caption{Schematic outlining steps in cavity solution. \textbf{1.} The initial parameter information consists of the probability distributions for the mechanistic parameters: $K_\alpha$, $m_i$ and $C_{i\alpha}$. We assume they can be described by their first and second moments.   \textbf{2.} The species dynamics $N_i(\sum_{\alpha}C_{i\alpha}R_\alpha-m_i)$ in eqs. (\ref{CRM_N}) are expressed as a factor graph.  \textbf{3.} Add  the "Cavity" species $0$ as the perturbation. \textbf{4.} Sum the resource abundance perturbations from the "Cavity" species 0 at  steady state and update the species abundance distribution to reflect the new steady state. \textbf{5.}  Employing the central limit theorem and the non-negativity constraint, the species distribution is expressed as a truncated normal distribution. \textbf{6.} Repeat \textbf{Step 2-4} for the resources.   \textbf{7.} The resource distribution is also expressed as a truncated normal distribution.  \textbf{8.} The self-consistency equations are obtained from the species and resource distributions. Figure taken from Ref.~\cite{cui2019effect}, \copyright~2020 American Physical Society.}
\label{cavity:scheme}
\end{figure*}

Just as we can use the cavity method to analyze the Lotka-Volterra model in high-dimensions, we can also adapt the cavity method to analyze consumer resource models (CRMs) in high dimensions. We will be interested in developing a statistical understanding of CRMs in the limit where the number of resources $M$ and the number of species $S$ in the regional species is large. In these lectures, we limit our analysis to MacArthur's original model ( eq. (\ref{eq:MacN}) and eq. (\ref{eq:MacR})).  For simplification, we set $w_\alpha=1$ and $r_\alpha=K_\alpha$ and get the following dynamics:
\begin{eqnarray}  
\frac{d N_i}{dt}  = N_{i} (\sum_{\beta} c_{i \beta} R_{\beta} - m_i)\label{CRM_N} , \\
\frac{d R_\alpha}{dt} =R_\alpha (K_{\alpha} -R_\alpha- \sum_{j}N_{j} c_{j\alpha}).\label{CRM_R} 
\end{eqnarray} 

The basic idea behind the cavity method is to derive  self-consistency equations by relating an ecosystem with $S$ species and $M$ resources to another ecosystem with  $S+1$ species and $M+1$ species. Unlike in the GLV model were we just had to one new species to the ecosystem, here it is important that we add a new species $N_0$ and new resource $R_0$ to the ecosystem (see Figure ~\ref{cavity:scheme}). We once again will work in the thermodynamic limit, $S, M \rightarrow \infty $, but with the ratio $S/M=\gamma^{-1}$ fixed. As for the GLV model, we further assume that the system is self-averaging and can be described using a ``replica symmetric'' ansatz. With these assumptions, just as in the GLV model, we will use perturbation theory to relate solutions in the $(S,M)$-ecosystem to the $(S+1,M+1)$-ecosystem. The logic behind the cavity solution of the MacArthur CRM is very similar to the derivations for the GLV model and for this reason this section is more condensed than the rest of these lectures.

The basic steps needed to derive the cavity solution are shown in Fig. \ref{cavity:scheme} and outlined below. {\bf Step 1}: the consumer preferences, $c_{i\alpha}$, are drawn from a Gaussian distribution with mean $\mu/M$ and variance $\sigma_c^2/M$.
They can be deposed into a deterministic and fluctuating component: 
\be
c_{i\alpha}=\mu/M + \sigma_c d_{i\alpha},
\label{Eq:cdecompose}
\ee
 where the fluctuating part $d_{i\alpha}$ obeys the statistics
\begin{eqnarray} \label{eq:d-mean}
\left< d_{i\alpha} \right> &=&0\\ \label{eq:d-var}
\left< d_{i\alpha} d_{j\beta} \right> &=&\frac{\delta_{ij}\delta_{\alpha\beta}}{M}.
\end{eqnarray}
We also assume that both the carrying capacity $K_\alpha$ and the minimum maintenance cost $m_i$ are independent Gaussian random variables with mean and covariance given by 
\begin{eqnarray}
\left< K_\alpha\right> &=& K\\
\text{Cov} (K_\alpha, K_\beta) &=& \delta_{\alpha\beta}\sigma^2_K\\
\left< m_i\right> &=& m\\
\text{Cov} (m_i, m_j) &=& \delta_{ij}\sigma^2_m.
\end{eqnarray}

We also define $\left< R\right>=\frac{1}{M}\sum_\beta R_\beta$ and $\left< N\right>=\frac{1}{S}\sum_j N_j$ to be the average resource and average species abundance, respectively. In terms of these quantities, we can re-write eq. (\ref{CRM_N}) and eq. (\ref{CRM_R}) as
\begin{equation}
\frac{d N_i}{dt} = N_i(\mu\left< R\right>-m+ \sum_{\beta}\sigma_c d_{i \beta} R_{\beta}-\delta m_i)\label{eq-r}
\end{equation}
\begin{equation}
\frac{d R_\alpha}{dt}= R_\alpha (  K \!+\! \delta K_\alpha \!\!-\!\!R_\alpha\!\!-\! \gamma^{-1} \mu\left< N\right>\!-\sum_j\sigma_c d_{j\alpha}N_j ) \label{eq-n} 
\end{equation}
where $\delta K_\alpha = K_\alpha  -K, \delta m_i = m_i-m$ and $\gamma^{-1} = S/M$. 

In {\bf Step 2}, we  express eq. (\ref{eq-n}) as a bipartite factor graph model for visualization. At {\bf step 3}, we add a ``cavity" species $N_0$ and  a ``cavity" resource $R_0$ into the ecosystem,
\begin{equation}
\frac{d N_0}{dt}= N_0(\mu\left< R\right>-m+ \sum_{\beta}\sigma_c d_{0\beta} R_{\beta}-\delta m_0)\label{dy-N0} 
\end{equation} 
\begin{equation} 
\frac{d R_0}{dt}= R_0 (K \!+\! \delta K_0\! -\!R_0 \! -\! \gamma^{-1} \mu\left< N\right>\!-\!\!\sum_j \sigma_c d_{j0}N_j ).  \label{dy-R0}
\end{equation} 

Adding new species and resource will perturb the original steady state. To characterize the perturbations, we introduce the following susceptibility matrices:
\begin{eqnarray}
\chi^R_{\alpha \beta}= \frac{\partial R_\alpha}{\partial K_\beta}, &\quad&
\chi^N_{i \alpha}=\frac{\partial N_i}{\partial K_\alpha},\label{eq:susceptibility1}\\
\nu^R_{\alpha i}=\frac{\partial R_\alpha }{\partial m_i}, &\quad &
\nu^N_{i j}=\frac{\partial N_i }{\partial m_j}\label{eq:susceptibility2}.
\end{eqnarray}
We can express the steady-state species and resource abundances in the $(S+1, M+1)$ system with a first-order Taylor expansion around the $(S,M)$ values. To do so, notice that
in the presence of the new species 
\begin{eqnarray}  
0 = N_{i} (\sum_{\beta} c_{i \beta} R_{\beta} - (m_i -c_{i0}R_0))\label{CRM_N_perturb} , \nonumber  \\
0 =R_\alpha ((K_{\alpha}-N_0c_{0 \alpha}) -R_\alpha- \sum_{j}N_{j} c_{j\alpha}) \label{CRM_R_perturb} 
\end{eqnarray} 
Since in eq. (\ref{Eq:cdecompose}), the contribution of the mean, $\frac{\mu}{M}\sim\mathcal{O}(\frac{1}{M})$, is much smaller than the fluctuation term, $\sigma_c d_{i0}\sim\mathcal{O}(\frac{1}{\sqrt{M}})$, to leading order the perturbations to $m_i$, and $K_\alpha$ in eqs. (\ref{eq-n}) and (\ref{eq-r}) are given by  $\sigma_c d_{i0}R_0$ and $\sigma_c d_{0\alpha}N_0$, respectively. Namely, in the presence of the new species (to leading order in $1/M$ or $1/S$) one has
\begin{equation}\label{eq:lcav}
N_i = N_{i /0} -\sigma_c\sum_{\beta /0} \chi^N_{i\beta} d_{0\beta}N_0 - \sigma_c\sum_{j /0}\nu^N_{ij} d_{j0}  R_0
\end{equation}
\begin{equation}\label{eq:Rcav}
R_\alpha = R_{\alpha /0} -\sigma_c\sum_{\beta /0} \chi^R_{\alpha  \beta} d_{0\beta}N_0 - \sigma_c\sum_{j /0}\nu^R_{\alpha j} d_{j0}  R_0
\end{equation}
Note  $\sum_{j /0}$ and $\sum_{\beta /0}$ mean the sums exclude the new species 0 and the new resource 0. These are just the analogue of Eqs.~\ref{eq:perturbNi} for the GLV model The next step is to plug eq. (\ref{eq:lcav}) and eq. (\ref{eq:Rcav}) into eq. (\ref{dy-N0}) and eq. (\ref{dy-R0}) and solve for the steady-state value of $N_0$ and $R_0$. 

\subsection{Self-consistency equations for species}
For the new ``cavity" species, the steady equation takes the form
\begin{eqnarray}
0&&=N_0 ( \mu\left< R\right>\!-\!m \!-\delta m_0\!-\!\sigma_c^2 N_0\!\!\sum_{\alpha/0, \beta/0}\chi^R_{\alpha\beta}d_{0\alpha}d_{0\beta}\\
&&\!\!-\sigma_c^2R_0\!\!\!\!\sum_{\beta /0, j /0}\nu^R_{\beta j}d_{0\beta}d_{j 0}\!+\!\!\!\sum_{\beta /0}\sigma_c d_{0\beta} R_{\beta/0} \!+\!\sigma_c d_{00} R_{0}).\nonumber
\label{Eq:n0mcrmeq}
\end{eqnarray}
Notice that each of the sums in this equation is the sum over a large number of uncorrelated random variables, and can therefore be well approximated by Gaussian random variables for large enough $M$ and $S$.   

As in the cavity solution of the GLV, we can replace the double sums over $d_{ j 0}d_{ 0 \beta}$ and $d_{0\alpha}d_{0\beta}$ by their expectation values using eq. (\ref{eq:d-mean}) and eq. (\ref{eq:d-var}):
\begin{eqnarray}\label{eq:chi=0}
\sum_{\beta /0, j /0}\nu^R_{\beta j}d_{0\beta}d_{j 0} =  \frac{1}{M}\sum_{\beta /0, j /0} \nu^R_{\beta j} \delta_{ j 0}\delta_{0 \beta}=0
\end{eqnarray}
\begin{equation}\label{eq:chi}
\sum_{\alpha/0 ,\beta /0}\chi^R_{\alpha\beta}d_{0\alpha}d_{0\beta} =  \frac{1}{M}\sum_{\alpha /0 ,\beta /0}\chi^R_{\alpha\beta} \delta_{\alpha\beta}=\chi 
\end{equation}
where $\chi=  \frac{1}{M}\sum_{\alpha /0 ,\beta /0}\chi^R_{\alpha\beta} \delta_{\alpha\beta}=\frac{1}{M}\text{Tr}(\chi^R_{\alpha\beta}) $ is the average susceptibility.  Substituting these expressions into eq. (\ref{Eq:n0mcrmeq}) , we obtain
\begin{eqnarray}
0&=& N_0 ( \mu\left< R\right>-m-\sigma_c^2\chi N_0 \nonumber\\
&+& \sum_{\beta /0}\sigma_c d_{0\beta} R_{\beta/0}-\delta m_0 )+ \mathcal{O}(M^{-1/2}).\label{sol-N0}
\end{eqnarray}

Instead of looking for a solution of $N_0$ for one realization of the random parameters, we can ask about the statistical distribution of the solution space for many different realizations of the random parameters. In this case, we know that  $\sum_{\beta /0}\sigma_c d_{0\beta} R_{\beta/0}$ can be well described by a Gaussian with mean
\be
\sigma_c \<\sum_{\beta /0} d_{0\beta} R_{\beta/0}\>=\sum_{\beta /0} \<d_{0\beta}\> R_{\beta/0}=0
\ee
and variance
\be
\<\left(\sum_{\beta /0}\sigma_c d_{0\beta} R_{\beta/0}\right)^2\>=\sigma_c^2 q_R,
\ee
where $q_R$ is the second moment of the resource distribution, 
$$q_R=\left< R^2 \right>=\frac{1}{M}\sum_\beta R_{\beta/0}^2.$$
Introducing an auxiliary random normal variable $z_N$ with zero mean and unit variance, this allows us to replace the sum below with a random variable as follows:
\begin{equation}\label{z_N}
\sum_{\beta /0}\sigma_c d_{0\beta} R_{\beta/0}-\delta m_0 = z_N\sqrt{\sigma_c^2 q_R+\sigma_m^2},\nonumber
 \end{equation}
 where we have used the fact that by assumption $\delta m_0$ is uncorrelated with $d_{0 \beta}$ and $R_{\beta /0}$. We can solve eq. (\ref{sol-N0}) in terms of the quantities just defined to get
\begin{equation}
 \mu\left< R\right>-m-\sigma_c^2\chi N_0+\sqrt{\sigma_c^2 q_R+\sigma_m^2}z_N= 0.  
 \end{equation}
Rearranging this to solve for $N_0$ and noting that $N_0 \ge 0$,  one gets
\begin{eqnarray}\label{N0}
N_0=\mathrm{max}\left( 0, \quad \frac{\mu\left< R\right>-m+\sqrt{\sigma_c^2 q_R+\sigma_m^2}z_N}{\sigma_c^2 \chi}\right)
 \end{eqnarray}
which is the formula for a  truncated Gaussian. In writing this solution, we have enforced the condition that, if possible, $N_0$ must be \emph{positive}. Ecologically, this is the statement that we are interested in the statistical description of uninvasible fixed points.

Combining eq. (\ref{N0}) and eq. (\ref{eq:wj}), by invoking self-averaging and replica symmetry we can easily write down the self-consistency equations for the fraction of non-zero species  as well as the moments of the species abundances at the steady state:
\begin{eqnarray}
\phi_N &=&\frac{S^*}{S}=  w_0\left(\frac{\mu\left< R\right>-m}{\sqrt{\sigma_c^2 q_R+\sigma_m^2}}\right)\label{phiN1}\\
 \left< N\right>&=&\left( \frac{\sqrt{\sigma_c^2 q_R+\sigma_m^2}}{\sigma_c^2 \chi}\right) w_1(\frac{ \mu\left< R\right>-m}{\sqrt{\sigma_c^2 q_R+\sigma_m^2}})\label{qN1}\\
 q_N&=&\left( \frac{\sqrt{\sigma_c^2 q_R+\sigma_m^2}}{\sigma_c^2 \chi}\right)^2 w_2(\frac{ \mu\left< R\right>-m}{\sqrt{\sigma_c^2 q_R+\sigma_m^2}})\label{qN2}
\end{eqnarray}

\subsection{Self-consistency equations for resource}
We now derive the equations for the steady-state of the resource dynamics. Inserting eq. (\ref{eq:Rcav}) into eq. (\ref{dy-R0})  gives:
\begin{eqnarray}
&&0=R_0 ( K\!-R_0-\gamma^{-1} \mu\left<N \right>+\!\sigma_c^2N_0\!\!\!\!\sum_{\beta/0, j/0}\!\!\!\!\chi^N_{j\beta}d_{j 0}d_{0\beta}\\
&&+\sigma_c^2R_0\!\!\!\!\!\sum_{i /0, j /0}\!\!\!\!\nu^N_{i j}d_{i 0}d_{j 0}\!-\!\sum_{j /0}\sigma_c d_{j0} N_{j/0}\!-\!\sigma_c d_{00} N_{0}+\!\delta K_0)\nonumber
\label{Eq:r0eqMCRM}
\end{eqnarray}
where $\gamma^{-1}=\frac{S}{M}$. 

To proceed, we note in the double sums above we can replace the sums over the $d_{i 0}d_{j 0}$ and $d_{j 0}d_{0\beta}$ by their expectation values using eq. (\ref{eq:d-mean}) and eq. (\ref{eq:d-var}) to get
\be
\sum_{i /0, j /0}\!\!\!\!\nu^N_{i j}d_{0i}d_{0 j}=\frac{S}{M}\times {1 \over S}\text{Tr}(\nu^N_{ij} ) =\gamma^{-1} \nu
\ee
and
\be
\sum_{\beta/0, j/0}\!\!\!\!\chi^N_{j\beta}d_{j 0}d_{0\beta}=0.
\ee
Furthermore, notice that $\sigma_c d_{00} N_{0}$ is of order $1/\sqrt{M}$ and can be ignored in
the thermodynamic limit. Finally, since we are interested in the statistical properties when we average over random parameters we can replace the sum over $N_{j/0}$ by Gaussian random variable of the form  
\be
\sum_{j /0}\sigma_c d_{j0} N_{j/0}\!+\!\delta K_0 = z_R\sqrt{\sigma_c^2\gamma^{-1}  q_N+\sigma_K^2},
\ee
where $z_R$ is unit normal random variable and $q_N=\left< N^2 \right>=\frac{1}{S}\sum_j N_j^2$. In writing this, we have once again used the fact that $\delta K_0$ is uncorrelated with $d_{j0}$ and $N_{j/0}$.

We can substitute these equations into eq. (\ref{Eq:r0eqMCRM}) and solve for $R_0$. If we further require that the resource are uninvasible (i.e. we always choose the positive solution for $R_0$ if it exists), then the distribution of $R_0$ is described by a truncated Gaussian of the form
\begin{eqnarray}
R_0&=&\mathrm{max}\left( 0, \quad \frac{K-\gamma^{-1} \mu\left< N\right>+\sqrt{\sigma_c^2\gamma^{-1}  q_N\!+\!\sigma_K^2} z_R}{1-\gamma^{-1} \sigma_c^2\nu}\right) \label{Mst1}.
\end{eqnarray}

This allows us to write down the self-consistency equations for the fraction of non-zero resources as well as the moments of their abundances at the steady state:
\begin{equation}
\phi_R\!=\!\frac{M^*}{M}=w_0(\frac{K-\gamma^{-1} \mu\left< N\right>}{\sqrt{\sigma_c^2\gamma^{-1}  q_N\!+\!\sigma_K^2}})\label{Mst4}
\end{equation}
\begin{equation}
\left< R\right>\!=\!\left(\!\frac{\sqrt{\sigma_c^2\gamma^{-1}  q_N+\sigma_K^2}}{1-\gamma^{-1} \sigma_c^2\nu}\right)\!w_1(\frac{K-\gamma^{-1} \mu\left< N\right>}{\sqrt{\sigma_c^2\gamma^{-1}  q_N\!+\!\sigma_K^2}})\label{Mst5}
\end{equation}
\begin{equation}
q_R\!=\!\left(\frac{\sqrt{\sigma_c^2\gamma^{-1}  q_N+\sigma_K^2}}{1-\gamma^{-1} \sigma_c^2\nu}\right)^2\!\!w_2(\frac{K-\gamma^{-1} \mu\left< N\right>}{\sqrt{\sigma_c^2\gamma^{-1}  q_N\!+\!\sigma_K^2}}).\label{Mst6}
\end{equation}

One can also use these equations to solve for the trace of the susceptibilities.
The susceptibilities are given by averaging $\nu^N_{ii}$ and $\chi^R_{\alpha\alpha}$,
\begin{eqnarray}
\chi &=& \left<\frac{\partial R_\alpha }{\partial K_\alpha}\right>=\frac{\partial \left<R\right> }{\partial K}=\frac{\phi_R}{1-\gamma^{-1} \sigma_c^2\nu}, \\
\nu &=& \left<\frac{\partial N_i }{\partial m_i}\right>=\frac{\partial \left<N\right> }{\partial m} =-\frac{\phi_N}{\sigma_c^2\chi}\label{Mst3}
\end{eqnarray}
Solving above two equations yields
\begin{eqnarray}
\chi =\phi_R-\gamma^{-1}  \phi_N, \quad \nu = - \frac{1}{\sigma_c^2}\frac{\phi_N}{\phi_R-\gamma^{-1} \phi_N} \label{eq:susceptibilities}.
\end{eqnarray}

Collectively Eqns~\ref{phiN1} -\ref{qN2},~\ref{Mst4}-~\ref{Mst6}, and ~\ref{eq:susceptibilities} define 8 self-consistency equations for the 8 quantities $\phi_N$, $\<N\>$, $\<N^2\>$, $\phi_R$, $\<R\>$, $\<R^2\>$, $\chi$, and $\nu$ that can be solved numerically.

\begin{figure}
\centering
\includegraphics[width=0.78\textwidth]{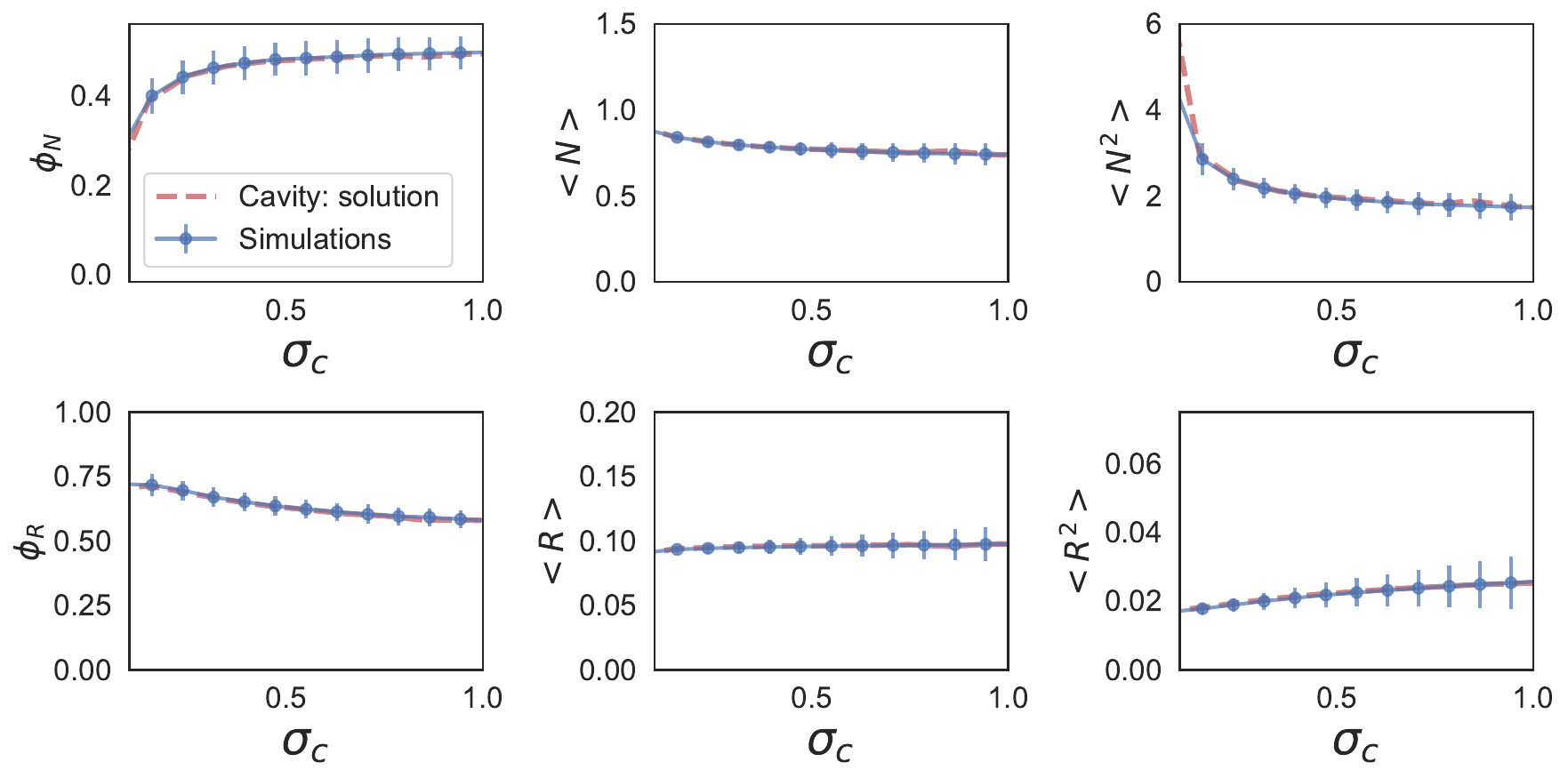}
\caption{Cavity predictions (obtained by solving the self-consistency equations derived in the text) compared to numerical simulations of MacArthur's consumer resource model with Gaussian consumer preferences. Panels show, as a function of $\sigma_c$, the fraction of species that survive, $\phi_N= \frac{S^*}{S}$, the fraction of resources that survive, $\phi_R= \frac{M^*}{M}$, and the first and second moments of the species and resource abundances. The error bar shows the standard deviation from $100$ numerical simulations with $M=S=100$ $\mu=1.$, $K=1.$, $\sigma_K=0.1$, $m=0.1$, $\sigma_m$=0.01. Simulations were run using the CVXPY package \cite{cvxpy_rewriting}.}
\label{cavity:solution}
\end{figure}

Fig. \ref{cavity:solution}
shows a comparison between the cavity solution and 1000 independent numerical simulations for various ecosystem properties such as the fraction of surviving species $S^*/S$, the fraction of surviving resources $M^*/M$, and the first and second moment of the species and resource distributions.  As can be seen in the figure,  our analytic expressions agree remarkably well over a large range of $\sigma_c$.

\begin{figure}
\centering
\includegraphics[width=0.78\textwidth]{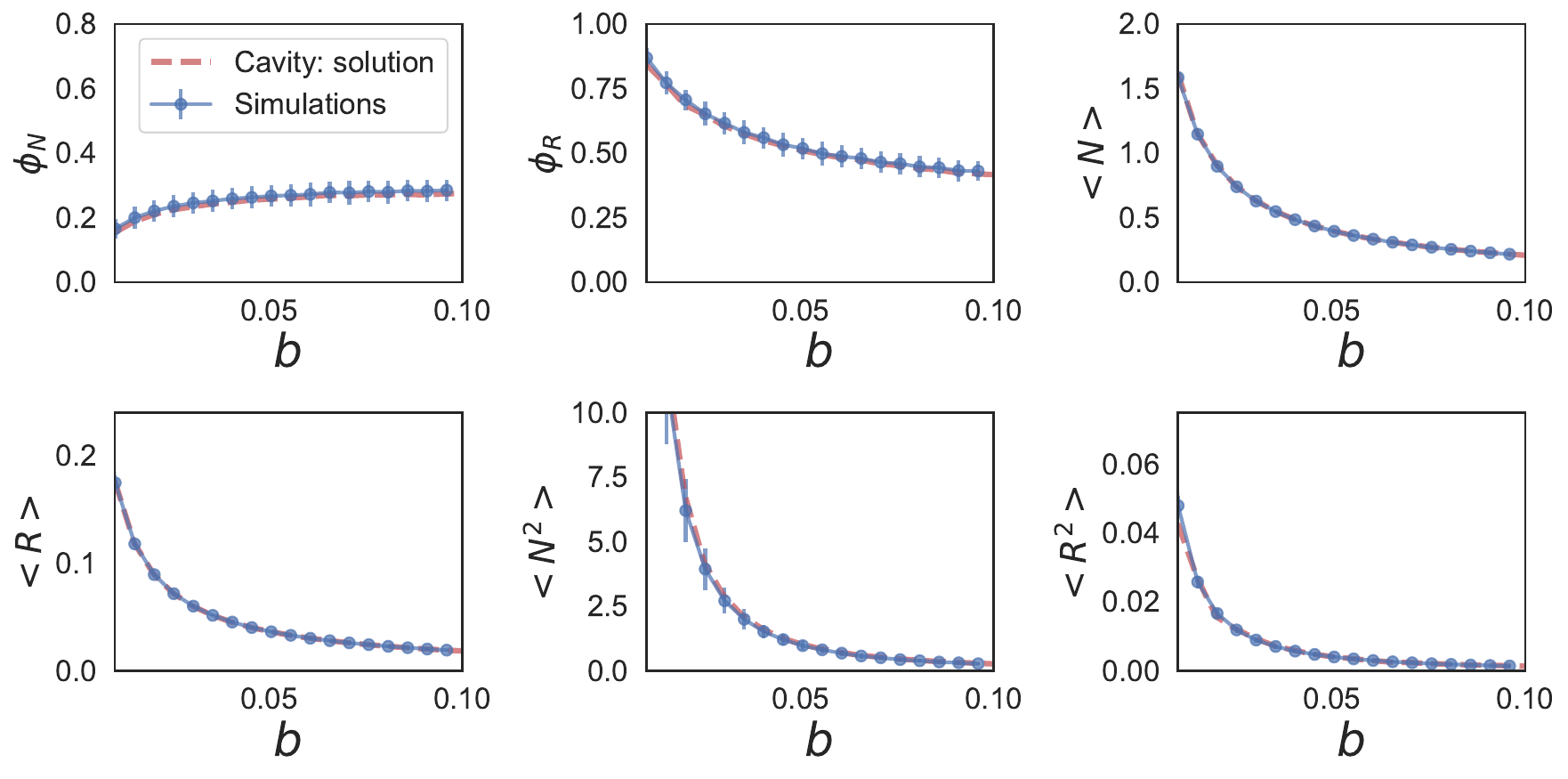}
\caption{Comparison between  cavity solutions and  simulations for MacArthur's consumer resource model. The consumer matrix is sampled from an uniform distribution $\mathcal{U}(0, b)$. The error bar shows the standard deviation from $500$ numerical simulations with $M=S=100$ $\mu=1.$, $K=1.$, $\sigma_K=0.1$, $m=0.1$, $\sigma_m=0.01$. Simulations were run using the CVXPY package \cite{cvxpy_rewriting}.}
\label{cavity:solution_uniform}
\end{figure}

\begin{figure}
\centering
\includegraphics[width=0.78\textwidth]{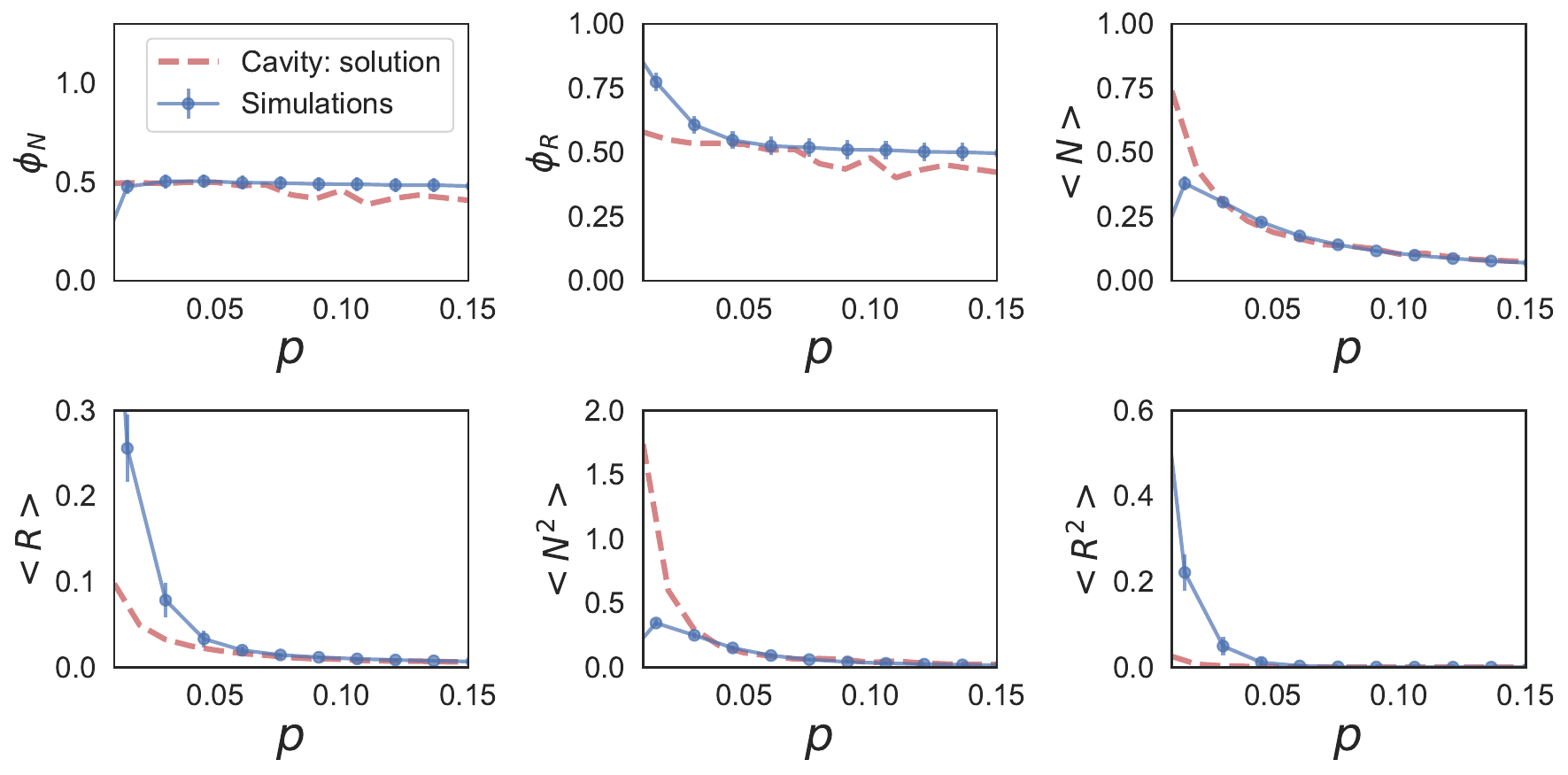}
\caption{Comparison between  cavity solutions and  simulations for MacArthur's consumer resource model. The consumer matrix is sampled from a Bernoulli distribution $\mathit{Bernoulli}(p)$. The error bar shows the standard deviation from $500$ numerical simulations with $M=S=100$ $\mu=1.$, $K=1.$, $\sigma_K=0.1$, $m=0.1$, $\sigma_m=0.01$. Simulations were run using the CVXPY package \cite{cvxpy_rewriting}.}
\label{cavity:solution_binomial}
\end{figure}

As a further check on our analytic solution, we ran simulations where the $c_{i \alpha}$ were drawn from different distributions. One pathology of choosing $c_{i \alpha}$ from a  Gaussian distribution is that when $\sigma_c$ is large,  many of consumption coefficients are negative. Another peculiarity of a Gaussian is that every species interacts with every other species  To test whether our cavity solution still describes ecosystems when $c_{i \alpha}$ are strictly positive or sparse,  we compare our cavity solution to simulations where the $c_{i \alpha}$  are drawn from either a Bernoulli or uniform distribution. The results are shown in Figs.~\ref{cavity:solution_uniform} and ~\ref{cavity:solution_binomial}.  As before, there is remarkable agreement between theoretical predictions and numerical simulations. The fact that our analytics also describe simulations from these alternative distributions highlights the incredible power of the central limit theorem and the cavity method. In fact, we expect that many properties  diverse ecosystems can be modeled using random ecosystems even when the consumer preferences have considerable structure, an idea explored using the cavity method and random matrix theory here \cite{ cui2021diverse}.
 
\section{Conclusions}

The goal of these lectures was to familiarize the reader with recent work at the intersection of statistical physics and community ecology. The bulk of the lectures focused on the two main models of community ecology: the generalized Lotka-Volterra model (GLV) and MacArthur's Consumer Resource Models (MCRM). We first presented these models in low-dimensions, where graphical methods could be used to develop intuition and understanding. We then discussed the deep connection between ecosystems with reciprocal interactions and constrained optimization before proceeding to analyze these same models in ``high-dimensions''  using methods from the statistical physics. 

We hope that these lectures can serve as a compact point of reference for those interested in community ecology. While many of the results presented here are scattered in the literature, it has been our experience there is a dearth of comprehensive pedagogical resources for beginners. It is our sincere hope that these lectures can fill this void.  As emphasized in the introduction, in no way are these lectures comprehensive. Instead, they should be viewed as a good starting point for delving into this rich and exciting research area.

\section{Acknowledgments}
We would like to thank the Mehta group (especially Zhijie ``Sarah'' Feng and Emmy Blumenthal) and our collaborators for countless conversations and insights.  We are extremely grateful to Arvind Murugan for his deep engagement with the work presented here, including critically reading an early version of these lectures and his many suggestions for improving the presentation. We would also like to thank Akshit Goyal for helpful comments on the presentation. This work was supported by NIH NIGMS R35GM119461 and a CZI Theory grant to PM.

%% file: 5_appendix.tex

\label{ch:appendix}

\section{Consumer resource models}
\label{sec:crm}

While all consumer resource models are broadly based on one idea - consumer growth depends on resource availability and consumer growth impacts resource availability, the  mathematical forms can vary significantly based on detailed assumptions made about the nature of consumer growth, resource consumption by consumers and resource renewal. Although in principle $g_i$, $q_{i\alpha}$ and $f_\alpha$ are chosen from an infinite variety of possible functions, in practice there are only a few forms that are commonly used. We will consider two possibilities for each, generating eight possible models, tabulated in Table \ref{tab:models}.
\begin{table}
\centering
	\begin{tabular}{c|l|l}
		\hline
		Function & Choice 1 & Choice 2\\
		\hline
		$g_i$ & Substitutable & Non-substitutable\\
		\hline
		$q_{i\alpha}$ & Mass action & Marginal benefit\\
		\hline
		$f_\alpha$ & Self-renewing & Externally supplied\\
		\hline
	\end{tabular}
\caption{Table of modeling choices}
\label{tab:models}
\end{table}

\subsection{Substitutable Resources}
For substitutable resources, the growth rate takes the form
\begin{align}
g_i(\mathbf{R}) = \sum_{\alpha} w_\alpha c_{i\alpha} R_\alpha - m_i
\end{align}
where $c_{i\alpha}$ is a set of foraging preferences, $w_\alpha$ represents the value of each resource in the ``common currency'' (e.g., energy), and $m_i$ is a minimal amount of consumption required to sustain growth.

\subsection{Nonsubstitutable Resources}
A general form for the growth rate with nonsubstitutable resources was given by Schreiber and Tobiason (2003):
\begin{align}
g_i(\mathbf{R}) = \left[\sum_{\alpha}w_\alpha (c_{i\alpha} R_\alpha)^n \right]^{\frac{1}{n}} - m_i
\end{align}
where the exponent $n$ can be any real number. Different values of $n$ correspond to different ecological categories of resources, as shown in Table \ref{tab:res} (cf. Koffel 2016). The weights $w_\alpha$ are typically set to 1, but are included so that this model reduces exactly to the above substitutable model when $n = 1$.
\begin{table}
\centering
	\begin{tabular}{c|c}
		\hline
		$n \to -\infty$ & Essential\\
		\hline
		$n < 0$ & Interactively essential\\
		\hline
		$0 < n < 1$ & Complementary\\
		\hline
		$n=1$ & Substitutable\\
		\hline
		$n > 1$ & Antagonistic\\
		\hline
		$n \to +\infty$ & Switching\\
		\hline
	\end{tabular}
\caption{Taxonomy of resource interactions}
\label{tab:res}
\end{table}

\subsection{Mass action}
The simplest and most common way of modeling the impact vectors is to assume mass action kinetics, where the depletion rate of resource $\alpha$ by consumer species $i$ is proportional to the rate of encounters between that resource and that species. This gives:
\begin{align}
q_{i\alpha}(\mathbf{R}) = -v_{i\alpha} R_\alpha
\end{align}
for some set of constants $v_{i\alpha}$ with units of [consumer density]$^{-1} \times$[time]$^{-1}$.

\subsection{Marginal benefit foraging}
(see \cite{letten2017linking}, p. 171)

The resource depletion rate in the mass action model is not related in any definite way to the growth rates. But in real systems, these two quantities are clearly related, because the growth is sustained by uptake of resources. One way of modeling this relationship is with marginal benefit foraging, where the relative foraging effort invested in each resource type is proportional to the partial derivative of the growth rate with respect to the (value-weighted) abundance of that resource. Instead of arbitrary constants $v_{i\alpha}$ determining the encounter rates between consumers and resources, the rates are now tied to the growth law by the following formula:
\begin{align}
q_{i\alpha}(\mathbf{R}) = - \frac{a_i}{w_\alpha} \frac{\partial g_i}{\partial R_\alpha} R_\alpha
\end{align}
where $a_i$ is a constant that sets the overall encounter rate, which we will always take to be equal to 1.

\subsection{Self-renewing resources}
In MacArthur's original consumer resource model, he considered biotic resources such as plants serving as food for herbivores, and modeled the intrinsic resource dynamics with a logistic law:
\begin{align}
f_\alpha(\mathbf{R}) = \frac{r_\alpha}{K_\alpha}R_\alpha(K_\alpha - R_\alpha)
\end{align}
where $K_\alpha$ is the carrying capacity and $r_\alpha$ is the low-density growth rate.

\subsection{Externally supplied resources}
When the resources are not alive, it is usually more physically reasonable to model them as diffusing in from some external source at some rate $\kappa_\alpha$, and decaying or diluting at a fixed rate $\omega_\alpha$, leading to the linear law
\begin{align}
f_\alpha(\mathbf{R}) = \kappa_\alpha - \omega_\alpha R_\alpha.
\end{align}

\subsection{Explicit equations for all models}
\subsubsection{Substitutable, Mass action, Self-renewing}
\begin{align}
\frac{dN_i}{dt} &= N_i \left[ \sum_{\beta} w_\beta c_{i\beta} R_\beta - m_i\right]\label{eq:MacNv}\\
\frac{dR_\alpha}{dt} &= \frac{r_\alpha}{K_\alpha} R_\alpha(K_\alpha - R_\alpha) - \sum_{j} v_{j\alpha} N_j R_\alpha.
\label{eq:MacRv}
\end{align}

\subsubsection{Substitutable, Mass action, Externally supplied}
\begin{align}
\frac{dN_i}{dt} &= N_i \left[ \sum_{\beta} w_\beta c_{i\beta} R_\beta - m_i\right]\\
\frac{dR_\alpha}{dt} &= \kappa_\alpha -\omega_\alpha R_\alpha - \sum_{j} v_{j\alpha} N_j R_\alpha.
\end{align}

\subsubsection{Nonsubstitutable, Mass action, Self-renewing}
\begin{align}
\frac{dN_i}{dt} &= N_i \left\{ \left[\sum_{\beta}w_\beta (c_{i\beta} R_\beta)^n \right]^{\frac{1}{n}} - m_i\right\}\\
\frac{dR_\alpha}{dt} &= \frac{r_\alpha}{K_\alpha} R_\alpha(K_\alpha - R_\alpha) - \sum_{j} v_{j\alpha} N_j R_\alpha
\end{align}

\subsubsection{Nonsubstitutable, Mass action, Externally supplied}
\begin{align}
\frac{dN_i}{dt} &= N_i \left\{ \left[\sum_{\beta}w_\beta (c_{i\beta} R_\beta)^n \right]^{\frac{1}{n}} - m_i\right\}\\
\frac{dR_\alpha}{dt} &=\kappa_\alpha -\omega_\alpha R_\alpha- \sum_{j} v_{j\alpha}N_j R_\alpha
\end{align}

\subsubsection{Substitutable, Marginal benefit, Self-renewing}
\begin{align}
\frac{dN_i}{dt} &= N_i \left[ \sum_{\beta} w_\beta c_{i\beta} R_\beta - m_i\right]\\
\frac{dR_\alpha}{dt} &= \frac{r_\alpha}{K_\alpha} R_\alpha(K_\alpha - R_\alpha) - \sum_{j} c_{j\alpha} N_j R_\alpha.
\end{align}

\subsubsection{Substitutable, Marginal benefit, Externally supplied}
\begin{align}
\frac{dN_i}{dt} &= N_i \left[ \sum_{\beta} w_\beta c_{i\beta} R_\beta - m_i\right]\\
\frac{dR_\alpha}{dt} &= \kappa_\alpha -\omega_\alpha R_\alpha - \sum_{j} c_{j\alpha} N_j R_\alpha.
\end{align}

\subsubsection{Nonsubstitutable, Marginal benefit, Self-renewing}
\begin{align}
\frac{dN_i}{dt} &= N_i \left\{ \left[\sum_{\beta}w_\beta (c_{i\beta} R_\beta)^n \right]^{\frac{1}{n}} - m_i\right\}\\
\frac{dR_\alpha}{dt} &= \frac{r_\alpha}{K_\alpha} R_\alpha(K_\alpha - R_\alpha) - \sum_{j} \left[\sum_{\alpha}w_\alpha (c_{j\alpha} R_\alpha)^n \right]^{\frac{1-n}{n}} c_{j\alpha}^n R_\alpha^{n-1} N_j R_\alpha.
\end{align}

\subsubsection{Nonsubstitutable, Marginal benefit, Externally supplied}
\begin{align}
\frac{dN_i}{dt} &= N_i \left\{ \left[\sum_{\beta}w_\beta (c_{i\beta} R_\beta)^n \right]^{\frac{1}{n}} - m_i\right\}\\
\frac{dR_\alpha}{dt} &= \kappa_\alpha -\omega_\alpha R_\alpha- \sum_{j} \left[\sum_{\alpha}w_\alpha (c_{j\alpha} R_\alpha)^n \right]^{\frac{1-n}{n}} c_{j\alpha}^n R_\alpha^{n-1} N_j R_\alpha.
\end{align}

\section{Self-limitation and stability}
\label{app:limit}

We can derive conditions for stability in the range of models described above. While the detailed analysis differs from case to case, qualitatively, the stability criterion is that each species must limit itself more than it limits others. The typical formulation of this in the niche theory framework is that each species consumes most of the resource that is most limiting to it. Here we generalize this statement to arbitrary numbers of species and resources.

Here, we first consider substitutable resources but different resource replenishment dynamics and impact vectors $q_{i\alpha}$. We show that non-substitutable resources can be handled by locally expanding the contours 

We fill focus on the basic example of MCRM with two resources and two species, but allowing an arbitrary impact vector $v_{i\alpha}R_\alpha$ that need not be related to $c_{i\alpha}$:
\begin{align}
    \frac{dN_i}{dt} &= N_i\left[ \sum_\alpha c_{i\alpha}R_\alpha - m_i\right]\\
    \frac{dR_\alpha}{dt} &= R_\alpha\left[\frac{r_\alpha}{K_\alpha}(K_\alpha - R_\alpha) - \sum_j N_j v_{j\alpha}\right].
\end{align}
The easiest thing to do is move to the ``fast resource limit'' (see below) and check the stability of the competition matrix. Following the derivation from the main text, and assuming no resources go extinct, we obtain:
\begin{align}
    \alpha_{ij} = \sum_\beta \frac{K_\beta}{r_\beta} c_{i\beta}v_{j\beta}.
\end{align}
For two species and two resources, we can explicitly compute the eigenvalues, finding:
\begin{align}
    &\lambda = \frac{1}{2}\bigg( \langle c_1|v_1\rangle + \langle c_2|v_2\rangle \nonumber\\
    &\pm \sqrt{(\langle c_1|v_1\rangle + \langle c_2|v_2\rangle)^2 - 4\frac{K_1 K_2}{r_1 r_2}|c||v|}\bigg)
\end{align}
where we have defined the inner product
\begin{align}
    \langle x|y\rangle \equiv \sum_\beta \frac{K_\beta}{r_\beta} x_\beta y_\beta,
\end{align}
and $|c|$ is the determinant of the matrix $c_{i\alpha}$. We see that all the eigenvalues are positive, implying stability, only when $|c|$ and $|v|$ have the same sign:
\begin{align}
    |c||v| > 0.
\end{align}
Writing out the determinants explicitly, we have:
\begin{align}
    |c| &= c_{11}c_{21}\left( \frac{c_{22}}{c_{21}} - \frac{c_{12}}{c_{11}}\right) = c_{11}c_{21}(\tan \theta^c_1 - \tan\theta^c_2)\\
    |v| &= v_{11}v_{21}\left( \frac{v_{22}}{v_{21}} - \frac{v_{12}}{v_{11}}\right) = v_{11}v_{21}(\tan \theta^v_1 - \tan\theta^v_2)
\end{align}
where the $\theta$'s denote the angle that the consumption vector and depletion vector of each species make with the resource 1 axis. The stability criterion thus implies that the species whose depletion vector $v_{i\alpha}$ makes a bigger angle with the resource 1 axis must also have the biggest angle for its consumption vector. If we say that the species with a larger $\theta^c_i$ is limited by resource 2, while the one with a smaller $\theta^c_i$ is limited by resource 1, then the equilibrium is stable whenever the depletion from each species is more biased towards its own limiting resource than the depletion from the other species is. But both depletion vectors could very well be extremely biased towards the same resource -- it's just the order that matters. And the magnitude of the impact doesn't matter at all, just the relative angle.